# Decoding dynamic brain patterns from evoked responses: A tutorial on multivariate pattern analysis applied to time-series neuroimaging data

Tijl Grootswagers[1*], Susan G. Wardle[1] and Thomas A. Carlson[1,2]

[1]Department of Cognitive Science and ARC Centre of Excellence in Cognition and its Disorders and Perception in Action Research Centre, Macquarie University, Australia

[2]School of Psychology, University of Sydney, Australia

*Corresponding Author:

Tijl Grootswagers

Department of Cognitive Science

Australian Hearing Hub

16 University Avenue

Macquarie University NSW 2109 Australia

Email: t.grootswagers@gmail.com

Acknowledgements. This research was supported by an Australian Research Council (ARC) Future Fellowship (FT120100816) and ARC Discovery project (DP160101300) awarded to T.A.C. S.G.W. is supported by an Australian NHMRC Early Career Fellowship (APP1072245). The authors declare no competing financial interests.

## Abstract

Multivariate pattern analysis (MVPA) or brain decoding methods have become standard practice in analysing fMRI data. Although decoding methods have been extensively applied in Brain Computing Interfaces (BCI), these methods have only recently been applied to time-series neuroimaging data such as MEG and EEG to address experimental questions in Cognitive Neuroscience. In a tutorial-style review, we describe a broad set of options to inform future time-series decoding studies from a Cognitive Neuroscience perspective. Using example MEG data, we illustrate the effects that different options in the decoding analysis pipeline can have on experimental results where the aim is to 'decode' different perceptual stimuli or cognitive states over time from dynamic brain activation patterns. We show that decisions made at both preprocessing (e.g., dimensionality reduction, subsampling, trial averaging) and decoding (e.g., classifier selection, cross-validation design) stages of the analysis can significantly affect the results. In addition to standard decoding, we describe extensions to MVPA for time-varying neuroimaging data including representational similarity analysis, temporal generalisation, and the interpretation of classifier weight maps. Finally, we outline important caveats in the design and interpretation of time-series decoding experiments.

## 1 Introduction

The application of 'brain decoding' methods to the analysis of fMRI data has been highly influential over the past 15 years in the field of Cognitive Neuroscience (Carlson, Schrater, & He, 2003; Cox & Savoy, 2003; Edelman, Grill-Spector, Kushnir, & Malach, 1998; Haxby et al., 2001; Kamitani & Tong, 2005). In addition to their increased sensitivity, the introduction of fMRI decoding methods offered the possibility to address questions about information processing in the human brain, which have complemented traditional univariate analysis techniques. Although decoding methods for time-series neuroimaging data such as MEG/EEG have been extensively applied in Brain Computing Interfaces (BCI; Curran & Stokes, 2003; Farwell & Donchin, 1988; Kübler, Kotchoubey, Kaiser, Wolpaw, & Birbaumer, 2001; K.-R. Müller et al., 2008; Vidal, 1973; Wolpaw, Birbaumer, McFarland, Pfurtscheller, & Vaughan, 2002), they have only recently been applied in Cognitive Neuroscience (Carlson, Hogendoorn, Kanai, Mesik, & Turret, 2011; Duncan et al., 2010; Schaefer, Farquhar, Blokland, Sadakata, & Desain, 2010).

The goal of this article is to provide a tutorial-style guide to the analysis of time-series neuroimaging data for Cognitive Neuroscience experiments. Although introductions to BCI exist (Blankertz, Lemm, Treder, Haufe, & Müller, 2011; Lemm, Blankertz, Dickhaus, & Müller, 2011), the aims of time-series decoding for Cognitive Neuroscience are distinct from those that drive the application of these methods in BCI, thus requiring a targeted introduction. While there are many reviews and tutorials for fMRI decoding (Cox & Savoy, 2003; Formisano, De Martino, & Valente, 2008; Haynes, 2015; Haynes & Rees, 2006; Mur, Bandettini, & Kriegeskorte, 2009; Norman, Polyn, Detre, & Haxby, 2006; Pereira, Mitchell, & Botvinick, 2009; Schwarzkopf & Rees, 2011), there are no existing tutorial introductions to decoding time-varying brain activity. Although the approaches are conceptually similar, there are important distinctions that stem from fundamental differences in the nature of the neuroimaging data between fMRI and MEG/EEG. In this paper, we provide a tutorial introduction using an example MEG data set. Although there are many possible analyses targeting time-series data (e.g., oscillatory (Jafarpour, Horner, Fuentemilla, Penny, & Duzel, 2013), or induced responses), we restrict the scope of this article to decoding information from evoked responses, with statistical inference at the group level on single time points or small time windows. As with most neuroimaging analysis techniques, the number of possible permutations for a given set of analysis decisions is very large, and the particular choice of analysis pipeline is guided by the experimental question at hand. Here we aim to provide a broad demonstration of how the analysis may be approached, rather than prescribing a particular analysis pipeline.

Early studies using time-resolved decoding methods have revealed significant potential for experimental investigation using this approach with MEG/EEG (see Section 1.1). However, compared to the popularity of decoding methods in fMRI, to date only a small number of studies have applied multivariate pattern analysis techniques to EEG or MEG. Accordingly, the aims of this article are to (a) Introduce the critical differences between decoding time-series (e.g., MEG/EEG) versus spatial (e.g., fMRI) neuroimaging data; (b) Illustrate the time-series decoding approach using a practical tutorial with example MEG data; (c) Demonstrate the effect that selecting different analysis parameters has on the results, and (d) Outline important caveats in the interpretation of time-series decoding studies. In sum, this article will provide a broad overview of available methods to inform future time-resolved decoding studies. This tutorial is presented in the context of MEG, however; the methods and analysis principles generalize to other time-varying brain recording techniques (e.g., ECoG, EEG, electrophysiological recordings.). As this review is targeted at providing a broad overview to a general audience, we avoid formal mathematical definitions and implementation details of the methods, and instead focus on the rationale behind the decoding approach as applied to time-series data.

### 1.1 Multivariate pattern analysis (MVPA) for MEG/EEG

The term 'multivariate pattern analysis' (or MVPA) encompasses a diverse set of methods for analysing neuroimaging data. The common element that unites these approaches is that they take into account the relationships between multiple variables, (e.g., voxels in fMRI, or channels in MEG/EEG), instead of treating them as independent and measuring relative activation strengths. The term 'decoding' refers to the prediction of a model from the data ('encoding' approaches do the reverse, predicting the data from the model, reviewed in Naselaris, Kay, Nishimoto, & Gallant, (2011), see also e.g., Ding & Simon, (2012) for an example of encoding models for MEG). The most common application of decoding in Cognitive Neuroscience is the use of machine learning classifiers (e.g., correlation classifiers (Haxby et al., 2001), or discriminant classifiers (Carlson et al., 2003; Cox & Savoy, 2003)) to identify patterns in neuroimaging data which correspond to the experimental task or stimulus. The most popular applications of MVPA are decoding (for recent reviews on fMRI decoding, see e.g., Haynes, 2015; Pereira et al., 2009), and more recently, Representational Similarity Analysis (RSA: Kriegeskorte & Kievit, (2013)). Within the broad category of MVPA analyses, the central focus of this article is on decoding methods applied to evoked responses, and the increasingly popular RSA framework (Section 5.2).

The decoding approach is illustrated in Figure 1 for a simple experimental design in which the subject viewed pictures of blue circles or red squares while their brain activity was recorded. The goal of the decoding analysis is to test whether we can predict if the subject was viewing a blue circle or red square, based on their patterns of brain activation. If the experimental stimuli can be successfully 'decoded' from the subject's patterns of brain activation, we can conclude that some information relevant to the experimental manipulation exists in the neuroimaging data. First, brain activation patterns in response to the different stimuli (or experimental conditions) are recorded using standard neuroimaging (e.g., MEG, fMRI, etc.) techniques (Figure 1A). The activation levels of the variables (e.g., voxels in fMRI, channels in MEG/EEG) in different experimental conditions are represented as complex patterns in high-dimensional space (each voxel, channel, or principal component is one dimension). For simplicity, in Figure 1B, these patterns are shown in two-dimensional space. Each point in the plot represents an experimental observation corresponding to the simultaneous activation level in two example voxels/channels in response to one of the experimental conditions (blue circles or red squares).

The first step in a decoding analysis involves training a classifier to associate brain activation patterns with the experimental conditions using a subset of the data (Figure 1C). In effect, during training the classifier finds the decision boundary in higher-dimensional space that best separates the patterns of brain activation corresponding to the two experimental categories into two distinct groups. As neuroimaging data is inherently noisy, this separation is not necessarily perfect (note the red square on the wrong side of the decision boundary in Figure 1C). Next, the trained classifier is used to *predict* the condition labels for new data that was not used for training the classifier (Figure 1D). The classifier predicts whether the new (unlabelled) data is more similar to the pattern of activation evoked by viewing a blue circle or a red square. If the classifier performs higher than that expected by chance (in this case 50% is the guessing rate as there are two stimuli), it provides evidence that the classifier can successfully generalize the learned associations to labelling new brain response patterns. Consequently, it is assumed that the patterns of brain activation contain information that distinguishes between the experimental conditions (i.e., the conditions blue circle/red square can be “decoded” from the neuroimaging data). Decoding accuracy can then be compared across brain regions (in fMRI), or time points (in MEG/EEG), in order to probe the location or time-course of information processing in the brain. This is achieved by repeating the classification multiple times for different data, that is, different time points in MEG/EEG (Figure 1E) for examining the time-course, or different brain regions in fMRI (Figure 1F) for examining the spatial distribution of information in the brain. Thus the main practical differences between decoding from MEG/EEG versus fMRI data lie in the methods used to obtain the patterns of information (Figure 1A, 1B), and the nature of the conclusions drawn from successful decoding performance (Figure 1E, 1F).

***** Figure 1 *****

Decoding time-series neuroimaging data is becoming increasingly popular. To date, most studies have applied the methods to understanding the temporal dynamics of the processing of visual stimuli and object categories. For example, time resolved decoding has been used to study the emergence of object representations at the category and exemplar level using MEG (Carlson, Tovar, Alink, & Kriegeskorte, 2013), EEG (Cauchoix, Barragan-Jason, Serre, & Barbeau, 2014), and neuronal recordings (Hung, Kreiman, Poggio, & DiCarlo, 2005; Meyers, Freedman, Kreiman, Miller, & Poggio, 2008; Zhang et al., 2011); how invariant object representations emerge over time (Carlson et al., 2011; Isik, Meyers, Leibo, & Poggio, 2014; Kaiser, Azzalini, & Peelen, 2016); and how objects are represented in other (e.g., written, or auditory) modalities (Chan, Halgren, Marinkovic, & Cash, 2010; Murphy et al., 2011; Simanova, van Gerven, Oostenveld, & Hagoort, 2014, 2010). Other studies have also used this approach to decode the orientation and spatial frequency of gratings from MEG (Cichy,

Ramirez, & Pantazis, 2015; Ramkumar, Jas, Pannasch, Hari, & Parkkonen, 2013; Wardle, Kriegeskorte, Grootswagers, Khaligh-Razavi, & Carlson, 2016), and to study decision making (Bode et al., 2012; Stokes et al., 2013), illusions (Hogendoorn, Verstraten, & Cavanagh, 2015), or working memory (van Gerven et al., 2013; Wolff, Ding, Myers, & Stokes, 2015). Notably, classifiers have been extensively applied to EEG (Guimaraes, Wong, Uy, Grosenick, & Suppes, 2007) for a different goal, as the low cost and portability of EEG is ideal for the development of brain computer interfaces (BCI). These applications use classifiers to predict brain states in order to operate computers or robots (Allison, Wolpaw, & Wolpaw, 2007; Hill et al., 2006; K. Müller, Anderson, & Birch, 2003; K.-R. Müller et al., 2008; Vidal, 1973, p. 2008). However, the goal of BCI is to achieve the maximum possible usability, i.e., optimal prediction accuracy, robust real-time classification, and generalizability. The performance measures of BCI systems are therefore often compared across studies (and in competitions; see e.g., Tangermann et al., 2012). This contrasts with decoding in neuroscience, where the goal is to understand brain processing by statistical inference on the availability of information (Hebart, Görgen, & Haynes, 2015), and accuracy differences between studies are generally not taken as meaningful.

Although the field is relatively new, there have already been several methodological extensions to standard decoding analysis applied to time-series neuroimaging data (Section 5). Following its application in fMRI, representational similarity analysis (Kriegeskorte & Kievit, 2013) has been used with MEG data to correlate the temporal structure of brain representations with behaviour (Redcay & Carlson, 2014; Wardle et al., 2016). RSA has also been used to link neuroimaging data from different modalities. For example, for object representations, the representational structure which appears early in the MEG data corresponds to representations in primary visual cortex measured with fMRI, whereas later stages instead reflect the representation in inferior temporal cortex (Cichy, Pantazis, & Oliva, 2014, 2016). A strength of time-series decoding is that the dynamic evolution of brain representations can be examined. One example of this is the temporal generalization approach (Section 5.1), which has been used in MEG to reveal that local and global responses to auditory novelty exhibit markedly different patterns of temporal generalisation (King, Gramfort, Schurger, Naccache, & Dehaene, 2014). Furthermore, insights into the spatiotemporal dynamics can also be gained by combining source reconstruction methods with the decoding approach (Sudre et al., 2012; van de Nieuwenhuijzen et al., 2013), or by comparing the interaction between subsets of sensors (e.g., Goddard, Carlson, Dermody, & Woolgar, 2016). Thus although relatively few time-series neuroimaging studies to date have applied decoding methods, these have already provided valuable insights, illustrating the rich potential for future applications.

Recently, several toolboxes have been developed that implement the methods described in the rest of this paper; The PyMVPA toolbox (Hanke, Halchenko, Sederberg, Hanson, et al., (2009); www.pymvpa.org) handles both fMRI and M/EEG data using the open-source Python language (Hanke, Halchenko, Sederberg, Olivetti, et al., 2009); MNE (Gramfort et al., (2013, 2014); http://martinos.org/mne) is a Python toolbox (and can be accessed in Matlab) designed for M/EEG analyses; the Neural Decoding Toolbox (Meyers, (2013); www.readout.info) is a Matlab toolbox created specifically for time-varying input; and the Matlab toolbox CoSMoMVPA (Oosterhof, Connolly, & Haxby, 2016; www.cosmomvpa.org) handles both fMRI and M/EEG, and was inspired by (and interfaces with) pyMVPA.

Decoding and other variants of MVPA are an alternative and complementary approach to univariate MEG/EEG analysis. This article will not cover univariate methods for MEG and EEG (which are well-established, see e.g., Cohen, 2014; Luck, 2005), and as always, the choice of analysis method must be guided by the experimental question. One of the central differences between univariate and multivariate methods is that the classifiers used in decoding approaches can use information that would not be detected when comparing the averaged signals in a univariate analysis (see Figure 2 for an illustration). This can lead to increased sensitivity for detecting differences between conditions (and on a single-trial basis). For example, decoding analysis can result in earlier detection of differences in the signals (Cauchoix, Arslan, Fize, & Serre, 2012; Cauchoix et al., 2014), and the differences found by classifiers can differ from those found in components (Ritchie, Tovar, & Carlson, 2015). Beyond sensitivity, the central distinction between univariate and MVPA analyses are the conceptual differences (activation-based versus information-based) in the experimental questions each approach is suited to addressing. We anticipate that time-series decoding approaches will continue to evolve alongside univariate methods, as has occurred with the adoption of decoding in fMRI, where both methods are used fruitfully.

***** Figure 2 *****

The main aim of this article is to describe a typical analysis pipeline for decoding time-series data in a tutorial format. The article is organized as follows; we begin by describing the experiment and the data-recording procedures used to obtain the example MEG data (Section 2). Next, we illustrate how the recordings are preprocessed using a combination of PCA, subsampling and averaging (Section 3). This is followed by the decoding analysis (Section 4). For all analysis stages we provide comparisons of how different choices made at each stage may affect the results. Following the decoding tutorial, in Section 5 we describe three extensions to the method: (1) temporal generalization (King & Dehaene, 2014), (2) representational similarity analysis (Kriegeskorte, Mur, & Bandettini, 2008), and (3) classifier weights projection (Haufe et al., 2014). Finally, we outline important caveats and limitations of the decoding approach in Section 6. See Figure 3 for an overview of the analysis pipeline and the structure of the paper, including the relevant section numbers.

***** Figure 3 *****

## 2 Description of experiment

In this tutorial, we use MEG data to illustrate the effect that different choices made at several analysis stages have on the decoding results. Object animacy has been shown to be a reliably decoded categorical distinction in studies using both fMRI (Downing, Chan, Peelen, Dodds, & Kanwisher, 2006; Kriegeskorte, Mur, Ruff, et al., 2008; Proklova, Kaiser, & Peelen, 2016; Sha et al., 2015) and MEG data (e.g., Carlson et al., 2013; Cichy et al., 2014). Here we use this robust paradigm as a basis for comparing the consequences of different analysis decisions in a decoding pipeline.

Twenty healthy volunteers (4 males) participated in the study with a mean age of 29.3 years (ranging between 24 and 35). Informed consent in writing was obtained from each participant prior to the experiment, and the study was conducted with the approval of the Macquarie University Human Research Ethics Committee. The stimuli were images of 48 visual object exemplars (24 animate and 24 inanimate) segmented and displayed on a phase-scrambled background (See Figure 4)[1]. Stimulus presentation was controlled by custom-written MATLAB (Natick, MA) scripts using functions from Psychtoolbox (Brainard, 1997; Kleiner et al., 2007; Pelli, 1997). The images were shown briefly for 66ms (at 9 degrees visual angle) followed by a fixation cross with a random inter-stimulus interval (ISI) between 1000 and 1200ms. Participants were instructed to categorize the stimulus as 'animate' or 'inanimate' as fast and accurate as possible, using a button press. The response button mapping alternated between 7-minute blocks, to avoid confounding the response with stimulus category (see

[1] The main study consisted of two conditions, stimuli in a clear or degraded state, however for the purpose of this article we only use the data for stimuli in the clear state (normal photographs of objects).

Section 6.1). This resulted in 32 trials per exemplar, 768 trials per category (animate/inanimate), and 1536 trials total per participant. All trials were included in the analysis, regardless of response, eye blinks or other movement artefacts.

***** Figure 4 *****

## 2.1. Data Collection

The MEG signal was continuously sampled at 1000Hz from 160 axial gradiometers[2] using a whole-head MEG system (Model PQ1160R-N2, KIT, Kanazawa, Japan) inside a magnetically shielded room (Fujihara Co. Ltd., Tokyo, Japan) while participants lay in a supine position. Recordings were filtered online with a high-pass filter of 0.03Hz and a low-pass filter of 200Hz. The recordings were imported into MATLAB using the Yokogawa MEG Reader Toolbox for MATLAB (YOKOGAWA Electric Corporation, 2011). The first step in the pipeline was to slice the data into epochs (i.e., trials), time-locked to a specific event. We extracted -100ms to 600ms of MEG data relative to the stimulus onset. The first 100ms of signal taken prior to trial onset serves as a sanity check for decoding accuracy (see Section 4.2).

## 2.2. Analysis Summary

The effect of different choices on the decoding results will be described by systematically varying one parameter relative to a set of fixed parameters. Three caveats of this approach are that (1) as these parameters are not independent, interactions between analysis decisions are likely, (2), the effects of these analysis decisions will vary between data sets, and (3), drawing conclusions on differences in decoding performances is only valid when the noise level is the same in all cases. Consequently, the following results should be interpreted as illustrative, rather than provide prescriptive analysis guidelines. All analysis code for the examples was written in Matlab (Natick, MA), using only standard functions unless otherwise specified. In order to illustrate the effects of different parameters on the results, they are consistently shown at the final stage plotted as a function of classifier accuracy over time. The default methods and fixed parameters are listed here for reference, and unless otherwise specified, the results in Figures 6-10 are obtained using this default pipeline:

- Preprocessing (Section 3)
    - Subsampling 200Hz
    - Averaging 4 trials
    - PCA retaining 99% of the variance
- Decoding (Section 4)
    - Naïve Bayes classifier
    - Leave-one-exemplar-out cross-validation

[2] Other MEG systems also include magnetometers, and there are possible differences in decodability from gradiometers and magnetometers (Kaiser, Azzalini, & Peelen, 2016).

The results are reported as time-varying decoding accuracy, i.e., higher accuracies reflect better decoding (prediction) of stimulus animacy from the MEG data. To assess whether accuracy was higher than chance, a Wilcoxon signed-rank test on the grand mean of decoding performance (*N*=20) was performed at each time point. The resulting *p*-values were corrected for multiple comparisons by controlling the false discovery rate (FDR, Benjamini & Hochberg, (1995)). Note that these statistics were chosen for their simplicity and ease of use, we discuss commonly used options for assessing classifier performance and statistics in Section 4.3.

Figure 5 shows the result of this default pipeline. As expected, before stimulus onset (-100 to 0ms), decoding performance is at chance (50%), confirming that there is no animacy information present in the signal. Then, approximately 80ms after stimulus presentation, the classifier's performance rises significantly above chance for almost the entire time window (to 600ms). Thus, at these time points, we are able to successfully decode from the MEG activation patterns whether the presented stimulus in a given trial was animate (e.g., parrot, dog, horse, etc.), or inanimate (e.g., banana, chair, tree, etc.). This indicates that the MEG signal contains information related to the animacy of the stimulus. The next sections will describe this pipeline in detail while comparing the effect of different analysis decisions.

***** Figure 5 *****

## 3. Preprocessing

Neuroimaging data is often noisy. The signals in imaging data are weak compared to, for example, environmental noise, baseline activity levels, or fluctuations caused by eye blinks or other movements. Therefore, a set of standard procedures is used to increase the signal-to-noise ratio. Furthermore, neuroimaging data are high-dimensional, and it is common practice to restrict the analysis to fewer dimensions. In MEG decoding, the dimensions of the data are generally reduced in the number of features (i.e., channels) that are input to the classifier. In addition, temporal smoothing is commonly applied. There are multiple ways to achieve these preprocessing steps, the most common are described in this section.

### 3.1 Data transformation and dimensionality reduction

A standard step in preprocessing is to reduce the dimensionality of the data. Some classifiers require more training samples than features; and others might overfit to noise in the data if provided with too many features (Bishop, 2006; De Martino et al., 2008; Misaki, Kim, Bandettini, & Kriegeskorte, 2010), or require longer computation time. Raw MEG recordings consist of many channels, typically 160 or more, and there is considerable redundant information, e.g., in adjacent channels. It is therefore common practice to reduce the dimensionality of the data by feature selection prior to decoding, which can be accomplished in multiple ways. One approach is to select the channels that are most informative (De Martino et al., 2008; Hanke, Halchenko, Sederberg, Hanson, et al., 2009). Isik et al., (2014) for example, by using an ANOVA significance test to select the MEG channels that contain significant stimulus-specific information.

Alternatively, one can use unsupervised, data-driven approaches such as Principal Component Analysis (PCA), which transforms the data into linearly uncorrelated components with the same number of feature dimensions, ordered by the amount of variance explained by each component (For a detailed introduction to PCA, see Jackson, (1991)). The use of PCA for MEG has a number of advantages: First, retaining only the components that account for most of the variance substantially reduces the dimensionality of the data. In the example data (160 channels), on average 48.16 (*SD*=7.05, range: 26 - 79) components accounted for 99% of the variance in the data. Secondly, PCA can separate out noise and artefacts such as eye blinks (see section 3.2) into their own components. These components can then be suppressed by the classifier because they do not contain class-specific information. Third, as the resulting PCA components are uncorrelated, it allows for using simpler (i.e., faster) classifiers that assume no feature covariance (e.g., Naïve Bayes, see Section 4.1).

Figure 6 illustrates the effect of the described dimensionality reduction methods on decoding performance for the example data. For this data set and classifier, PCA yields much better performance compared to using the raw channels (cf. Isik et al., 2014). Note that these differences are classifier dependent (as shown in Section 4.1). Here, the PCA transformation was computed on the training data, and applied on the test data, separately for each time point, and separately for each training fold. Alternatively, one could compute *one* transformation for the whole time-series, and/or do this on *all* data before the cross-validation process. However, this is only viable if the goal of the analysis is statistical inference (Hebart et al., 2015), as it could result in more optimistic decoding accuracies that would not generalize to new data[3].

***** Figure 6 *****

An alternative method is to transform the sensor-level data into activations in virtual source space. Instead of decoding channel-level activations, source reconstruction (e.g., beamformer (Van Veen, Van Drongelen, Yuchtman, & Suzuki, 1997), or minimum norm estimate (Hämäläinen & Ilmoniemi, 1994)) can be applied during preprocessing. Classification is then performed in source space rather than channel space (Sandberg et al., 2013; Sudre et al., 2012; van de Nieuwenhuijzen et al., 2013). Using source space for decoding has the potential to improve classification accuracies (Sandberg et al., 2013; van de Nieuwenhuijzen et al., 2013), as source reconstruction algorithms can ignore channel-level noise. Inferences about the spatial origin of the decoded discrimination can be made by restricting the classifier to considering signals from pre-determined regions of interest (e.g., Sudre et al., 2012), or by using the complete source space reconstruction and projecting the classifier weights (see Section 5.3) into source space (e.g., van de Nieuwenhuijzen et al., 2013). The second approach relies on interpreting classifier weights, and therefore, the reliability of the sources depends not only on the reconstruction quality, but also on decoding performance (see Section 5.3). Source reconstruction methods are still developing, and reconstruction accuracies are likely to improve in the future, making source space decoding an attractive option. However, as source space decoding has not been widely used to date, we will not cover it in the rest of this tutorial.

## 3.2 Improving signal to noise

[3] Note that when comparing PCA performed inside the cross-validation loop on separate time points with PCA performed before the cross-validation on all time points, we did not find any difference in classifier accuracy (data not shown), but this may not hold for different data sets.

MEG data is generally sampled at high frequencies (e.g., 1000Hz), and a common strategy to improve signal-to-noise (the strength of the signal compared to the strength of the background noise) is by collapsing data over time. The two main approaches are to classify on more than one time point using a sliding window (e.g., Ramkumar et al., 2013), or down-sample the data to lower frequencies (see Figure 7). The difference between the methods is that when using a sliding window, the classifier has access to all time points in the window (the number of features is increased), while in subsampling, it receives the average (the number of features at each time point stays the same). For the example data, subsampling has a small effect on decoding performance, but also benefits the analysis by reducing the computation time for the decoding analysis as there are fewer time points to classify. The sliding window approach also improves performance, but the benefit is marginal especially considering that the computation time increases significantly with larger sliding windows as the classifier is still trained and tested at each time point. The optimal parameters will depend on the particular data set and desired temporal resolution. An important caveat for both approaches is that estimates of both decoding onset and the time of peak decoding are affected by the choice of subsampling or sliding window. When using a sliding time window, the last time bin in the window should be used for determining the onset (as in Figure 7), to avoid shifting the onset forward in time. It is recommended to apply a low-pass filter before resampling (e.g., subsampling using the *decimate* function in MATLAB) as subsampling can cause aliasing. Low-pass filtering, however, can cause an artefact whereby significant decoding emerges even when no signal exists in the original data (Vanrullen, 2011). For the example data, we subsampled by a factor of 5 to obtain a sampling rate of 200Hz.

*** **Figure 7** ***

Another source of noise originates from artefacts. Eye blinks, eye movements, heartbeats, and muscle movement can cause significant artefacts. Typically, in classical M/EEG analyses trials containing such artefacts are manually inspected and excluded from the analysis, or independent component analysis is used to separate out these artefacts into their own components, which are then removed manually or automatically (Mognon, Jovicich, Bruzzone, & Buiatti, 2011). Experiments can also be designed in a way to reduce the number of artefacts, for example by instructing participants to blink in response to a particular stimulus that is not part of the analysis (Cichy et al., 2014). We did not perform any artefact rejection on our data, and found classification performance to be well above chance, but this can vary across data sets. As classifiers have the capacity to learn to ignore bad channels or supress noise during training, artefact correction is likely less critical in decoding analyses. However, note that if artefacts are confounded with a condition (e.g., if more eye movements occurred in one condition than the other due to some property of the stimulus), this would make the artefacts a potential source of discrimination information for the classifier. If this is the case, it would not be possible to determine whether the classifier was decoding the experimental condition, or the correlated difference in artefacts (see also Section 6.1).

Increased signal-to-noise can also be achieved by averaging trials belonging to the same exemplar before decoding (Isik et al., 2014). Averaging increases general decoding performance and makes signatures (e.g., onsets, maxima or minima) more pronounced. This effect is shown in Figure 8, where different numbers of trials (belonging to the same exemplar) are averaged. Interestingly, the first onset of decoding is similar regardless of the number of trials that are averaged. The greatest increase in performance (in our example data) is observed when averaging 4 trials. Averaging more trials does not increase decoding performance by the same factor, suggesting that here 4 trials is a good trade-off between signal-to-noise, and trials per exemplar. The trade-off to consider when selecting the number of trials to average is that reducing the trials per exemplar (e.g., averaging 32 trials here produces only one trial per exemplar) typically increases the variance in (within-subject) classifier performance. Alternatively, when not enough trials are available, the trials used for training the classifier could be sampled with replacement (bootstrapped). The optimal number of trials to average will differ for different data (e.g., in Isik et al., (2014), averaging 10 trials was used). Note that trial averaging does not affect model testing (e.g., RSA, Section 5.2), as relative decoding performance is scaled similarly between exemplars or time points.

***** Figure 8 *****

## 4. Decoding

Decoding analysis is performed on the preprocessed data. To summarize, in preprocessing the raw MEG signal is sliced into epochs from -100 to 600ms relative to stimulus onset, then down-sampled to 200Hz. Groups of 4 single trials are averaged to boost signal-to-noise, resulting in 8 pseudo-trials for each object exemplar. These preprocessed pseudo-trials are the input to the classifier in the decoding analysis.

In order to decode the class information (animacy) from the MEG data, a pattern classifier (see Section 4.1) is trained to distinguish between two classes of stimuli (animate and inanimate objects). The classifier's ability to generalize this distinction to new data is assessed using cross-validation (see Section 4.2). If the classifier's performance after cross-validation is significantly above chance, this indicates that the MEG patterns contain class-specific information, and we conclude that the class can be decoded from the MEG data. In time-resolved MEG decoding studies, this process is repeated on all time points in the data. Then, for example, one can examine when the peak in decoding performance occurs, i.e., at what time point the information in the signal allows for the best class distinction. Another feature often used is the onset of significant decoding performance, to determine the earliest time that class-specific information becomes available. These signatures can then be compared across experimental conditions.

### 4.1 Classifiers

There are numerous types of classifiers, which originate from the machine learning literature. Classifier choice has the potential to influence experimental results, as different classifiers make different assumptions about the data. In addition, the goal of classification in machine learning is high predication accuracy, which drives the development of increasingly sophisticated classifier algorithms. In contrast, prediction is not the main goal of decoding in neuroscience, and classifier choice instead favours simplicity and ease of interpretation over optimizing prediction accuracies. Therefore, for brain decoding studies, linear classifiers are generally preferred, as they are simpler in nature, making interpretation less complex (Misaki et al., 2010; K. Müller et al., 2003; Schwarzkopf & Rees, 2011). The default classifiers used in fMRI decoding are typically linear support-vector machines (SVM), or, to a lesser extent, correlation classifiers. However, fMRI data typically has many features/dimensions. SVM is generally better than other classifiers when dealing with many features and is therefore a popular choice. In comparison to fMRI data, time-series data often has fewer features (e.g., our example MEG data set uses only ~50 components following PCA). Consequently, it is possible that there are differences in the suitability of different classifiers for fMRI versus time-series decoding analysis. Here we compare the performance of SVM, correlation classifiers, and two common alternatives (Linear Discriminant Analysis (LDA) and Gaussian Naïve Bayes (GNB)) on the example MEG data (Figure 9), using their built-in Matlab implementations (and default parameters). Notably, LDA, GNB and SVM have the best overall performance. Taking the complexity of the classifier into account, which affects the computational requirements and given that classification is generally repeated many times (e.g., on multiple time points), this argues in favour of the discriminant classifiers (GNB and LDA), which are faster to train than SVM. Interestingly, despite their relative popularity in fMRI, the correlation classifiers did not perform as well on our data. However, Isik et al., (2014) reported correlation classifier performance for their MEG data on par with other classifiers. This difference could be due to many factors, for example, different choices in the preprocessing pipeline or experimental design. To illustrate that classifier performance depends on preprocessing, we tested the same classifiers using different preprocessing decisions. For example, Figure 9B shows that not performing PCA has a large effect on GNB performance, but a smaller effect on the performance of LDA and SVM. These dependencies highlight the difficulty in attempting to make universal recommendations for decoding analyses. Furthermore, each classifier has a number of parameters that may be optimised, however, most neuroscience studies use standard classifier implementations.

***** Figure 9 *****

### 4.2 Cross-validation

An essential step in decoding analysis is cross-validation: this provides an evaluation of classifier generalization performance. In standard *k*-fold cross-validation, the data is divided into *k* subsets (i.e., folds), where each subset contains a balanced amount of trials from each class (e.g., animate and inanimate exemplars in our example experiment). The classifier is trained using all-but-one subsets (the training set). Next, the trained classifier is used to predict the class of the trials from the remaining subset (the test set). This process is repeated for all subsets, and the average classifier performance across all folds is reported. This method makes maximal use of the available data, as all trials are used for testing the classifier. Note that in fMRI decoding the sets are often based on experimental runs (leave-one-*run*-out cross validation), as the trials within each run are not independent (e.g., due to the slow hemodynamic response). In MEG decoding, individual trials are generally assumed to be independent (Oosterhof et al., 2016), and trials are randomly assigned to train and test sets. The theoretical optimal performance is obtained by leave-one-*trial*-out cross-validation, where the classifier is trained on all-but-one trial. It is however computationally more intensive, especially with many trials (which is typically the case in MEG).

As with other analysis decisions, the most appropriate implementation of cross-validation is guided by the experimental design. Standard *k*-fold cross-validation assigns individual trials to training and testing sets. Depending on the research question, this may produce a confound in the class distinction that the classifier learns from the training data. For example, for decoding animacy, standard cross-validation would entail that trials belonging to the same exemplar (e.g., 'car') are assigned to both training and test sets. Consequently, it may be possible for the classifier to learn to distinguish the classes based on the activation patterns evoked by visual properties of specific exemplars. This makes it unclear whether the classification boundary is based on animacy or visual features. To avoid this, when decoding categories composed of many exemplars; we recommend leave-one-*exemplar*-out cross-validation (see Carlson et al., 2013), where all trials belonging to one exemplar (e.g., car) are assigned to the test set and the classifier is trained on the data from the other exemplars (e.g., 'dog' and 'chair'). This is repeated for all exemplars (i.e., every exemplar is assigned to the test set once).

Figure 10 shows decoding accuracy for different forms of cross validation, including an invalid analysis without cross validation. Note that without cross validation, classifier performance is above chance *prior* to stimulus onset. This nonsensical result arises from the test data being used to train the classifier, violating the constraint of independence. Time-resolved decoding methods have a convenient built-in check for this: above chance decoding performance before stimulus onset suggests an error exists in either the preprocessing or cross validation stages. In our data, 10-fold and leave-one-trial-out cross validation yielded very similar results, suggesting that the optimal split is data-specific. Further, by comparing performance between traditional cross validation (e.g., *k*-fold) and leave-one-*exemplar*-out, it is possible to estimate to what degree classifier performance is driven by individual stimulus properties (e.g., low-level visual properties of the exemplar images). The difference between *k*-fold and leave-one-exemplar-out cross validation is observed early in the time-series (consistent with the timing of early visual feature processing), and is reduced later in the time course (Figure 10). Taken together, a valid form of cross-validation with independent training and test data is essential. Although there are several ways of splitting up the data into training and test sets, the particular version of cross-validation implemented must be compatible with the research question.

***** Figure 10 *****

### 4.3 Evaluation of classifier performance and group-level statistical testing

Statistical evaluation of decoding analyses is a complex issue, and there is not yet consensus on the optimal approach (Allefeld, Görgen, & Haynes, 2015; Nichols & Holmes, 2002; Noirhomme et al., 2014; Schreiber & Krekelberg, 2013; Stelzer, Chen, & Turner, 2013). The statistical approach used in our example analysis is common in the literature (e.g., Carlson, Tovar, et al., 2013; Ritchie et al., 2015) and was chosen for its simplicity; however, there are several alternative methods that are also valid. For example, we report classifier performance as accuracy (percent correct). Accuracy is a less appropriate measure when dealing with unbalanced data (more trials exist for one class than for the other), as a trained classifier could exploit the uneven distribution and achieve high accuracy simply by predicting the more frequent class. For unbalanced data, a measure of performance that is unaffected by class bias such as D-prime is more appropriate. Alternatively, 'balanced accuracy' includes the mean of the accuracies for each class and thus is also unaffected by any class imbalance in the data.

Several options exist for assessing whether classifier performance is significantly above chance. The non-parametric Wilcoxon signed-rank test (Wilcoxon, 1945) was used in our example (see also Carlson, Tovar, et al., 2013; Ritchie et al., 2015), as it makes minimal assumptions about the distribution of the data. Alternatively, the Student's t-test is also commonly used (but see Allefeld, Görgen, & Haynes, 2016). Another popular alternative is the permutation test, which entails repeatedly shuffling the data and recomputing classifier performance on the shuffled data to obtain a null-distribution, which is then compared against observed classifier performance on the original set to assess statistical significance (see e.g., Cichy et al., 2014; Isik et al., 2014; Kaiser et al., 2016). Permutation tests are especially useful when no assumptions about the null-distribution can be made (e.g., in the case of biased classifiers or unbalanced data), but they take much longer to run (e.g., repeating the analysis ~10,000 times).

Importantly, as is the case in fMRI analyses, time-series neuroimaging analyses also require addressing the problem of multiple comparisons (Bennett, Baird, Miller, & Wolford, 2011; Bennett, Wolford, & Miller, 2009; Nichols, 2012; Pantazis, Nichols, Baillet, & Leahy, 2005) as typically multiple tests are conducted across different time points. The FDR adjustment used in our example analysis is straightforward, but a limitation is that it does not incorporate the relation between time points (Chumbley & Friston, 2009). Alternatively, cluster-based multiple-comparison correction involves testing whether *clusters* of time points show above-chance decoding and therefore can result in increased sensitivity to smaller, but more sustained effects (Mensen & Khatami, 2013; Nichols, 2012; Oosterhof et al., 2016; Smith & Nichols, 2009).

## 5. Additional analyses

In the sections above, we illustrated the standard approach to decoding time-series neuroimaging data. Here we outline three extensions for decoding analysis. The first is temporal cross-decoding (Section 5.1), which tests the degree to which activation patterns in response to the experimental conditions are sustained or evolve over time. The second is the RSA framework (Section 5.2), which facilitates the testing of models of the structure of decodable information over time. Finally, we outline a method that involves projection of the classifier weights in order to determine the spatial source of the signal driving the classifier in sensor-space (Section 5.3).

### 5.1 The temporal generalization method

An advantage of time-series decoding is that it has the potential to reveal the temporal evolution of brain activation patterns, rather than providing a single, static estimate of decodability for a stimulus or task. One method is to train a classifier on a particular time point, and then test its decoding performance on *different* time points. This form of cross-decoding reveals to what degree the activation patterns for a particular stimulus or task evolve. Classifiers effectively carve up multidimensional space in order to distinguish between the experimental conditions, thus when a classifier which is trained on one time point can successfully predict class-labels for data at other time points, it suggests that the structure of the multidimensional space is similar across time. Conversely, if cross-decoding is unsuccessful across two time points, it suggests that the multidimensional space has changed sufficiently for the boundary between classes determined at one time point to be no longer meaningful by the second time point. Beyond temporal characterisation of the decoding results, this method has the potential utility to test cognitive models which make theoretical predictions about the generalizability of representations (see also Figure 4 in King & Dehaene, 2014). For example, the temporal generalization of classifiers can be tested between two completely separate datasets. Isik et al., (2014) tested the temporal generalization performance of a classifier that was trained on stimuli that were presented foveally, and then tested on peripherally presented stimuli. Similarly, Kaiser et al., (2016) used this method to distinguish category-specific responses from shape-specific responses.

Figure 11A shows cross-validated temporal cross-decoding performed on the example MEG data. The diagonal in this figure is analogous to the standard one-dimensional time-series decoding plot (e.g., Figures 5-10). Significant points (shown in Figure 11B) off the diagonal indicate that the classifier, when trained on data from time point A, can generalize to data from time point B. The generalization accuracy normally drops off systematically away from the diagonal. In this case, classifier performance generalizes well for neighbouring time points (red region on the diagonal) as expected, and additionally, to some extent between 150-200 and 300-500ms, indicating that the MEG activation patterns are similar in these windows.

***** Figure 11 *****

### 5.2 Representational Similarity Analysis (RSA)

Standard decoding analysis reveals whether class-specific information is present in the neuroimaging signal. Approaches such as cross-decoding (e.g., temporal generalisation) can begin to probe the underlying representational structure of the information in the brain activation patterns used by the classifier. RSA takes this concept further, and provides a framework for testing hypotheses about the structure of this information (Kriegeskorte, Mur,

& Bandettini, 2008). RSA is based on the assumption that stimuli with more similar neural representations are more difficult to decode. Conversely, stimuli with more distinct representations are expected to be easier to decode. Thus the central idea is that representational similarity can be indexed by the degree of decodability. By comparing the decodability of all possible pair-wise combinations of stimuli, a representational dissimilarity matrix (RDM) is calculated. That is, for each pair of stimuli, the distance between their activation patterns is computed using one of several distance metrics (e.g., correlation between the activation patterns, or difference in classifier performance (Walther et al., 2015).

An example RDM is shown in Figure 12A, in which each cell in the matrix corresponds to the dissimilarity of two of the object stimuli in the MEG animacy experiment. For data with high temporal resolution such as MEG, a series of RDMs can be created for each time point, and used to investigate the temporal dynamics of representations over time. The time-varying RDMs in Figure 12A are constructed by decoding all pairwise stimuli using the same pipeline (using 2-fold cross-validation, as leave-one-exemplar out is not possible when decoding between 2 exemplars), thus one square in the RDM represents the decoding accuracy for classifying between one pair. Following calculation of the RDM (either time-varying or static) from the empirical data, the empirical RDM can be compared to model RDMs that make specific predictions about the relative decodability of the stimulus pairs. In RSA studies to date, model RDMs have been constructed from predictions based on a wide range of sources: including behavioural results, computational models, stimulus properties, or neuroimaging data from a complementary imaging method such as fMRI (e.g., Carlson, Simmons, Kriegeskorte, & Slevc, 2013; Cichy et al., 2014, 2016; Kriegeskorte, Mur, Ruff, et al., 2008; Redcay & Carlson, 2014; Wardle et al., 2016).

Figure 12 shows the results of RSA model evaluation for the example MEG data. For each time point, the empirical RDMs (Figure 12A) are correlated with three theoretical models (Figure 12B); a model of stimulus animacy, a model that distinguishes artificial versus natural stimuli, and a control model based on the visual similarity of the exemplar's silhouettes (which correlates well with early stimulus discriminability, see e.g., Carlson et al., 2011; Redcay & Carlson, 2014). Each of these models predicts the relative (dis)similarity of the MEG activation patterns for each exemplar pair based on their specific stimulus features. The extent of the correlation between the model and empirical MEG RDMs is interpreted as reflecting the degree to which the 'representational structure' characterised by each model exists in the brain activation patterns. The results in Figure 12C are plotted as the correlation between the three model RDMs with the MEG RDM over time. The Animacy model (blue line) has a better fit to the MEG data than the Natural model (orange line), and both models have a better fit than the Silhouette model (yellow line) later in the time series. The Silhouette model has the best fit early in the time series, which is expected as it represents early visual features. This suggests that animacy is a relatively good predictor of the similarity of the MEG activation patterns for the exemplar pairs: object pairs from the same category (e.g., both animate) are more difficult to decode than object pairs from different categories (e.g., one animate and one inanimate). Within the RSA framework, this is interpreted as evidence that animacy is a key organising principle in the representational structure of the object exemplars.

Despite its strengths, a current limitation of the RSA approach is that valid statistical comparison of different candidate models is difficult (Kriegeskorte & Kievit, 2013; Thirion, Pedregosa, Eickenberg, & Varoquaux, 2015). A recent development proposes evaluating model performance by comparing it to the highest possible performance given the noise in the data, called the 'noise ceiling' (Nili et al., 2014). When applied to MEG data, the performance of various models relative to the noise ceiling (computed from the empirical data as described in Nili et al., 2014) can be evaluated over time, as shown in Figure 12C. Despite the present limitations in directly comparing different models, RSA is a useful tool for investigating the structure of the decodable signal in neuroimaging data, which will undoubtedly continue to evolve in its sophistication and utility. For a more detailed introduction, see Kriegeskorte and Kievit (2013), Kriegeskorte, Mur, and Bandettini (2008), and Nili et al., (2014).

***** Figure 12 *****

### 5.3 Weight projection

Following successful classification of experimental conditions, it is sometimes of interest to examine the extent to which different voxels (fMRI) or sensors (MEG/EEG) drive classifier performance. During standard classification analysis, each feature (e.g., MEG sensors) is assigned a weight corresponding to the degree to which its output is used by the classifier to maximize class separation. Therefore, it is tempting to use the raw weight as an index of the degree to which sensors contained class-specific information. However, this is not straightforward, as higher raw weights do not directly imply more class-specific information than lower weights. Similarly, a non-zero weight does not imply that there is class-specific information in a sensor (for a full explanation, proof, and example scenarios, see Haufe et al., 2014). This is because sensors may be assigned a non-zero weight not only because they contain class-specific information, but also when their output is useful to the classifier in suppressing noise or distractor signals (e.g., eyeblinks or heartbeats). An elegant solution to this issue was recently introduced by Haufe et al., (2014) and has been applied to MEG decoding (Wardle et al., 2016). This consists of transforming the classifier weights back into activation patterns. Following this transformation, the reconstructed patterns are interpretable (i.e., non-zero values imply class-specific information) and can be projected onto the sensors. It is import to note however, that the reliability of the patterns depends on the quality of the weights. That is, if decoding performance is low, weights are likely sub-optimal, and reconstructed activation patterns have to be interpreted with caution (Haufe et al., 2014).

Here we summarize this transformation for MEG data, and plot the results in Figure 13. First, the classifier weights (we used LDA instead of GNB in this example as this method only applies to classifiers that consider the feature covariance) are transformed into activation patterns by multiplying them with the covariance in the data: $A = \text{cov}(X)*w$; where $X$ is the $NxM$ matrix of MEG data with $N$ trials and $M$ features (channels), and $w$ is a classifier weight vector of length $M$. $A$ is the resulting vector of length $M$ containing the reconstructed activation patterns (i.e., the transformed classifier weights). For display purposes, the reconstructed activation patterns can be projected onto the scalp location of the channels. Figure 13B shows the result for the example MEG data at four time points (using the FieldTrip toolbox for MATLAB: Oostenveld, Fries, Maris, & Schoffelen, 2010); here the results are scaled by the inverse of the source covariance ($\text{A}*\text{cov}(\text{X}*\text{w})^{-1}$) to allow for comparison across time points. Note that this method cannot be directly used if multiple time points are used for classification (e.g., the sliding window approach described in Section 3.2). The uncorrected (raw) weight projections are shown for comparison in Figure 13A. We can now observe that for the activation patterns in Figure 13B, the information source is located approximately around the occipital lobes (back sensors) at 100ms, and later around the temporal lobes (side sensors) at 300ms, as expected from the visual processing hierarchy. Notably, this pattern is not as easily identifiable in the raw weight topographies shown in Figure 13A. For an in-depth explanation (with examples) of the weights interpretation problem and its solution, see Haufe et al., (2014).

***** Figure 13 *****

## 6. General discussion

Time-series decoding methods provide a valuable tool for investigating the temporal dynamics and organization of information processing in the human brain. In the previous sections we outlined an example decoding analysis pipeline for time-series neuroimaging data, illustrated effects of different methods and parameters (and their interactions), and introduced extensions of the method such as temporal generalisation (5.1), RSA (5.2), and weights projection (5.3). In the final section, we discuss some important aspects to consider when performing these analyses and interpreting the results. One of the central issues concerns the interpretation of classifier accuracy. Classifiers are extremely sensitive and will exploit all possible information in the data. This means that careful experimental design and interpretation of the results is required in order to draw meaningful conclusions from decoding studies (see e.g., Carlson & Wardle, 2015; de-Wit, Alexander, Ekroll, & Wagemans, 2016; Naselaris & Kay, 2015). The next section outlines a number of such pitfalls to avoid in the implementation of time-series decoding methods.

### 6.1 Common pitfalls

The first caveat applies to all studies using classifiers and is well-described in the literature (Kriegeskorte, Lindquist, Nichols, Poldrack, & Vul, 2010; Kriegeskorte, Simmons, Bellgowan, & Baker, 2009; Pereira et al., 2009). It is important that the classifier has no access to class-specific information about the data contained in the test set, as this will artificially inflate classifier performance. This analysis confound is referred to as 'double dipping', and was demonstrated in the analysis without cross-validation in Figure 10 (Section 4.2). One advantage of time-series decoding is that in most cases, data obtained before stimulus onset serves as a first check. If classifier accuracy is above chance before stimulus onset, it indicates possible contamination from double dipping.

A second caveat specific to time-series decoding is that caution is required when interpreting (differences in) onsets of significant decoding. The time at which decoding is first significant for an experimental condition is determined by the underlying strength of the signal. For example, when the strength of peak decoding differs between two conditions (e.g., one is much easier to decode than the other), this will also affect the relative onset of decoding. This is illustrated in Figure 14. Three simulated data sets were constructed to have the same decoding onset (50ms) and peak latency of decoding (100ms), but different signal strengths (see Figure 14A). To evaluate how signal strength influences decoding onset, Gaussian noise was added to each data set and significance testing was conducted to find the onset of decoding (signed-rank test across time points, FDR corrected). The outcome of the simulation is plotted in Figure 14B. Note that even though these simulated data sets were constructed to have an identical 'true' onset of decoding, the onset of significant decoding is earlier for the set with a strong signal and much later for the set with the weak signal. This underscores the ambiguity in interpreting onset differences: it cannot be assumed that an earlier decoding onset reflects a true onset difference in the availability of decodable information between conditions. Isik et al., (2014) addresses this issue by using less data for the condition that had higher peak decoding, and by equalizing the peaks across conditions before determining decoding onset.

***** Figure 14 *****

Third, as noted earlier, filtering the signal can smear out information over time. An extreme example (using a step function) is illustrated in Figure 15, using simulated data with a signal occurring at 50ms. To demonstrate the effect of filtering, Gaussian noise was added to the signal, and low-pass filters were applied with different cut-off frequencies using the *ft_preproc_lowpassfilter* function (using the default Butterworth 4$^{th}$ order two-pass IIR filter) from the FieldTrip toolbox (Oostenveld et al., 2010). The result of lowering the cut-off frequency is increased signal distortion. Applying a 30Hz low-pass filter resulted in a signal that was significantly different from zero 40ms earlier in the time-series, compared to the simulated 'true' onset at 50ms. However, the effect is substantially reduced by applying much higher filter cut-offs, e.g., 200Hz. Therefore, interpretations based on the timing of decoding signatures relative to the stimulus should be avoided when using filters with a low cut-off frequency (Vanrullen, 2011).

***** Figure 15 *****

Finally, decoding studies require careful experimental design to avoid confounds in the classifier analysis. The considerations vital to designing decoding studies are not necessarily the same as that for univariate analysis. Accordingly, care must be taken when re-analysing data not originally intended for a decoding analysis. The high sensitivity of classifiers means that if there are any differences between classes other than the intended manipulations, it is likely that the classifier will exploit this information, making it easy to introduce experimental confounds. An example is the effect of the subject's behavioural responses. In our example MEG experiment, the response buttons (to respond 'animate' and 'inanimate') were switched every block. If response mapping were uniform across blocks, response would be confounded with stimulus category, as a left button response would always correspond to 'animate', and right for 'inanimate'. The physical pressing of the button would generate corresponding brain signals, for example in motor areas, and this would provide a signal in the whole-brain MEG data that would correlate perfectly with the class conditions. In this case, it would be unclear whether the classifier decoded the intended experimental manipulation of 'animacy' or simply the subject's motor responses. Alternatively, a classifier may distinguish between two conditions or categories of stimuli based on a confounding factor that co-varies with class membership (e.g., differential attention to two conditions, leading to greater overall signal for one class) rather than the manipulation (e.g., difference in visual features or task difficulty) intended by the experimental design.

Further, even with carefully controlled designs, the interpretation of decoding studies must be executed with caution. Decoding studies may conclude that condition A is decodable from condition B; however, the source of decodable information usually remains elusive (Carlson & Wardle, 2015; Naselaris & Kay, 2015). One notable example of this is the current debate surrounding the source of orientation decoding in fMRI (e.g., Alink, Krugliak, Walther, & Kriegeskorte, 2013; Carlson, 2014; Carlson & Wardle, 2015; Clifford & Mannion, 2015; Freeman, Ziemba, Heeger, Simoncelli, & Movshon, 2013; Freeman, Brouwer, Heeger, & Merriam, 2011; Kamitani & Tong, 2005; Mannion, McDonald, & Clifford, 2009; Pratte, Sy, Swisher, & Tong, 2016). Despite a decade of orientation decoding in early visual cortex with fMRI, it is still debated whether any information at the sub-voxel level (e.g., within-voxel biases in orientation-specific columnar responses) contributes to the decodable signal (Op de Beeck, 2010). The interpretation of the source of decodable signals in neuroimaging remains one of the central challenges facing the application of MVPA techniques to advancing our understanding of information processing in the human brain (de-Wit et al., 2016).

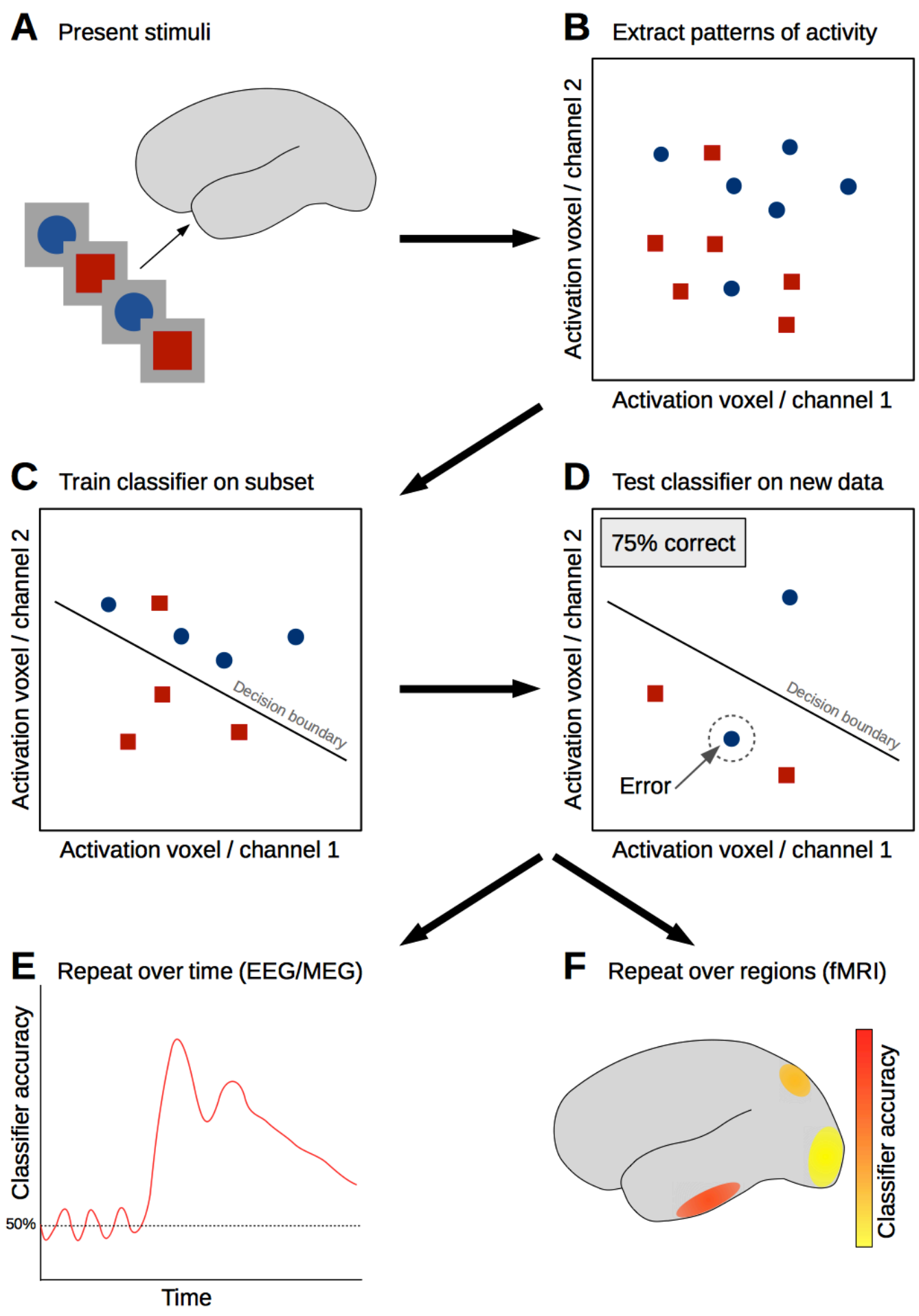

**Figure 1.** The general decoding approach. **A.** Brain responses to stimuli (e.g. blue circles and red squares) are recorded with standard neuroimaging techniques. **B.** Patterns of activation evoked by the two stimulus conditions (red square and blue circle) are represented in multiple dimensions (channels in EEG/MEG, or voxels in fMRI); here only two dimensions are illustrated for simplicity. **C.** A classifier is trained on a subset of the neuroimaging data, with the aim of distinguishing a reliable difference in the complex brain activation patterns associated with each stimulus class. **D.** The performance of the classifier in distinguishing between the stimulus classes is evaluated by testing its predictions on independent neuroimaging data (not used in training) to obtain a measure of decoding accuracy. **E,F.** Steps B-D may then be repeated for different time points (when using EEG/MEG) to study the temporal evolution of the decodable signal, or repeated for different brain areas (in fMRI) to examine the spatial location of the decodable information.

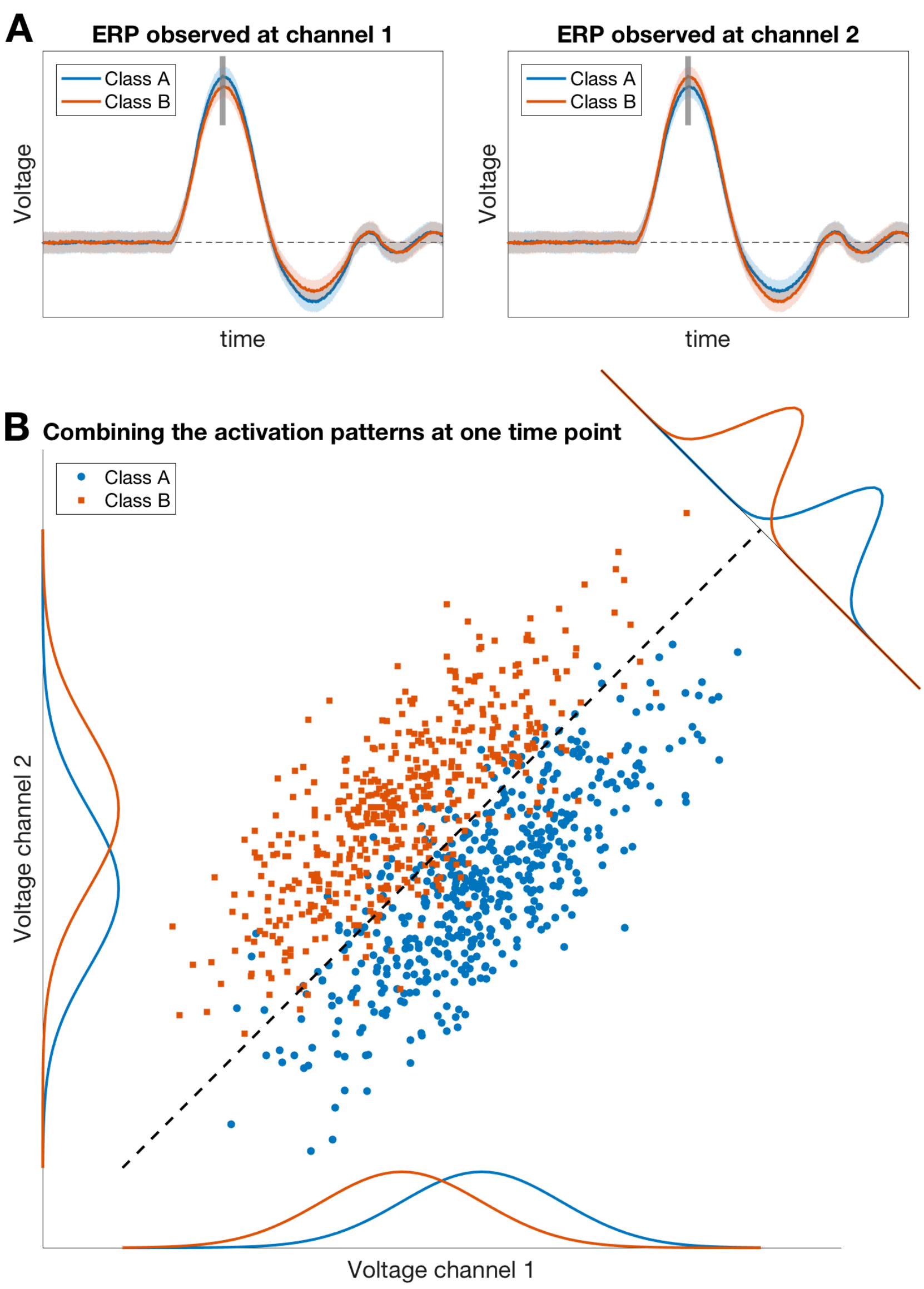

**Figure 2.** An illustration of how multivariate analysis can result in increased sensitivity compared to univariate analysis. **A.** Example average event-related potentials (ERPs) in response to two stimuli (class A and class B) are shown in two channels (left and right panels). The responses to the two classes in the individual channels overlap substantially, and potentially non-significant in a univariate analysis. **B.** The same responses represented as points in two-dimensional space, showing the activation in the two channels at one time point (i.e., location of the vertical grey bar in the ERP plots). When combining the information from both channels as in a decoding analysis, it is possible to define a boundary (dashed line) separating the two classes (distributions plotted orthogonal to the dashed line).

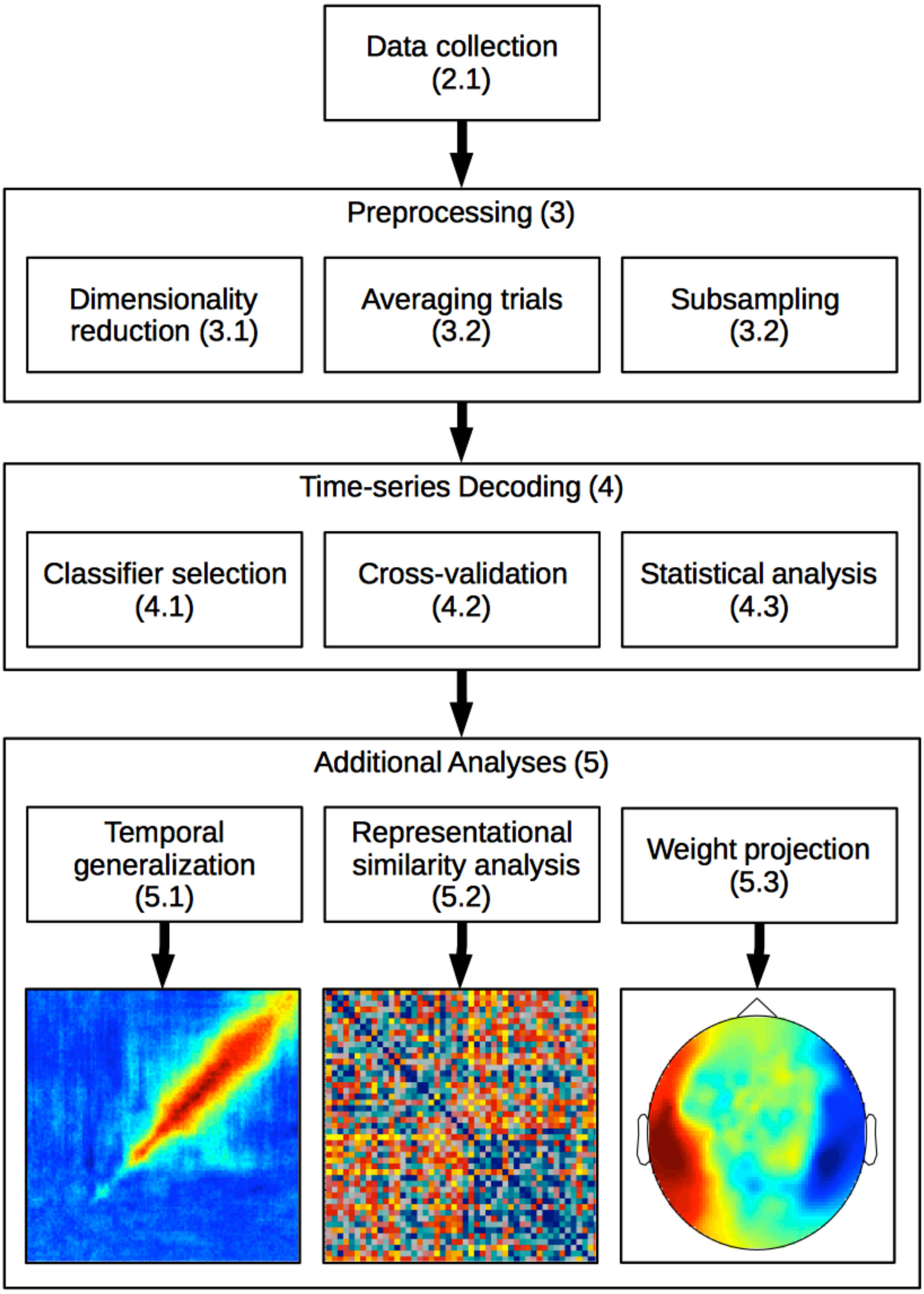

**Figure 3.** A schematic overview of a typical analysis pipeline. Refer to the relevant sections in the article for further details (numbers in the figure indicate section numbers). This overview illustrates a general pipeline for decoding studies. The practical differences between decoding with MEG/EEG data versus fMRI data arise in both the preprocessing and analysis stages.

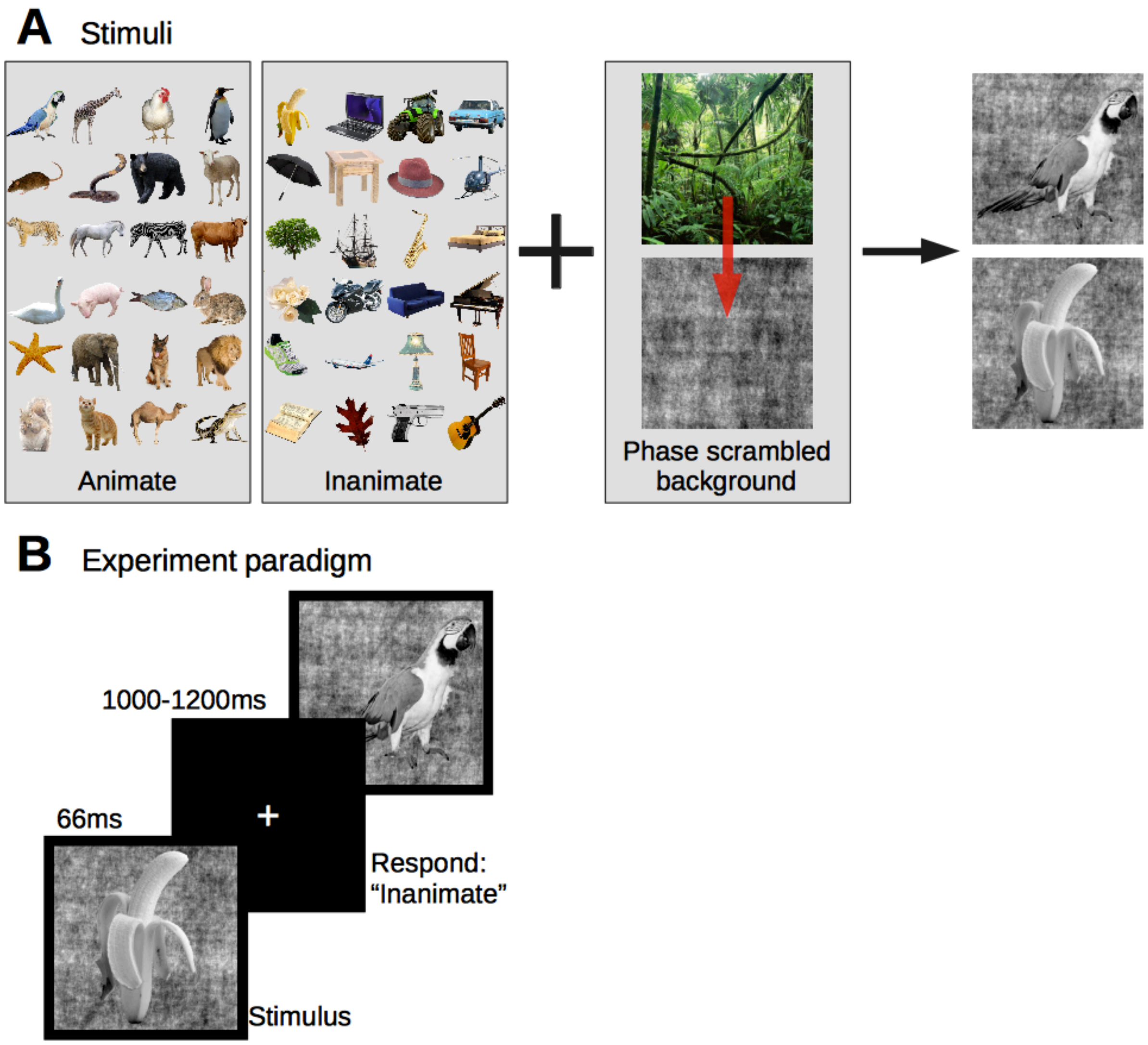

**Figure 4.** Illustration of the experimental design. **A.** The stimuli consisted of 24 animate and 24 inanimate visual objects, converted to grey-scale and overlayed on a phase-scrambled natural image background. **B.** Stimuli were presented in random order for 66ms followed by a random ISI between 1000 and 1200ms. Participants categorized the animacy of the stimulus during the ISI with a button press.

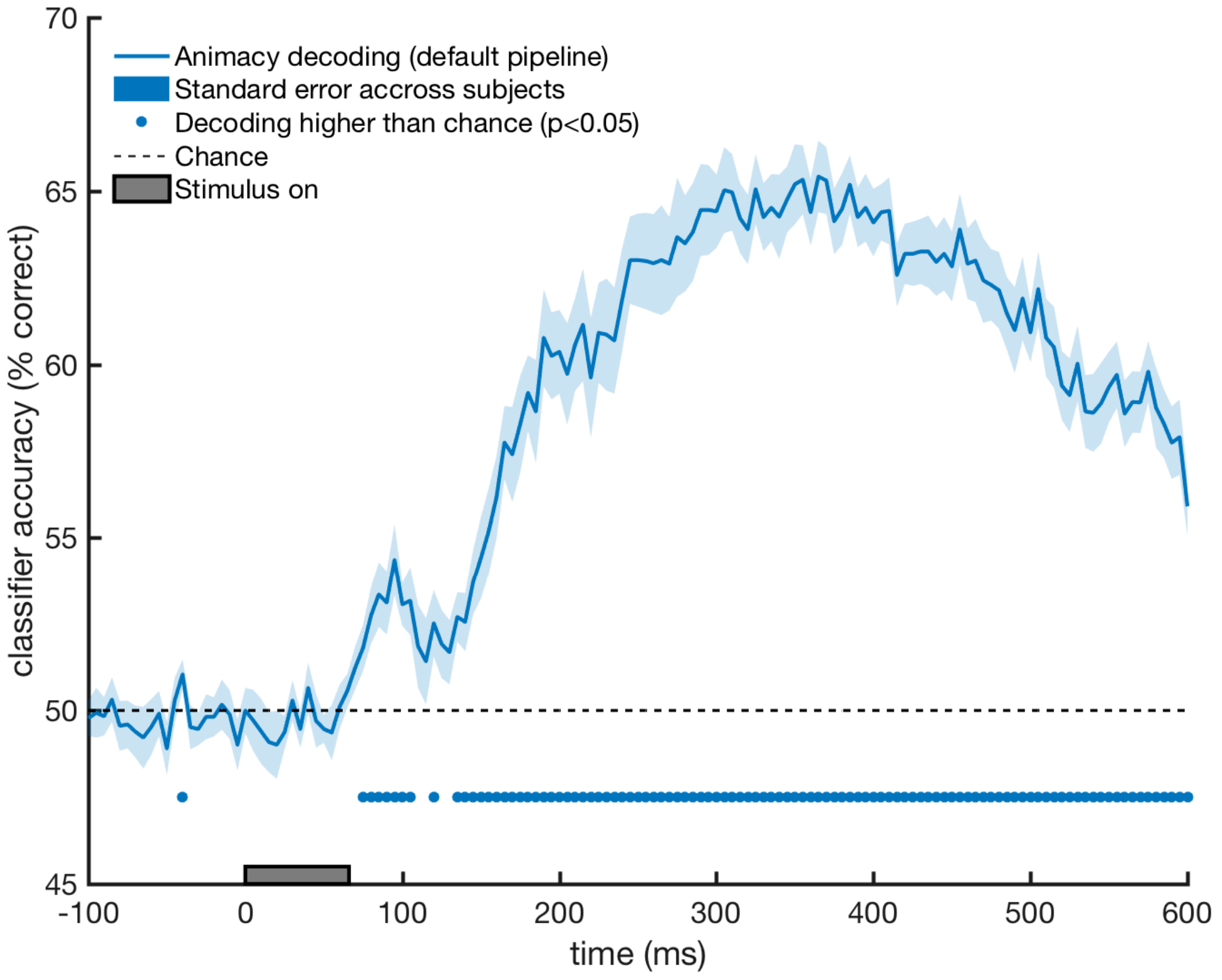

**Figure 5.** Decoding animacy from MEG data using the default analysis pipeline. Classifier accuracy (percent correct averaged across subjects) is shown as a function of time relative to stimulus onset at 0ms. The dashed line marks chance classification accuracy at 50%. The shaded area is the standard error across subjects. Discs above the x-axis indicate the time points where decoding performance is significantly higher than chance.

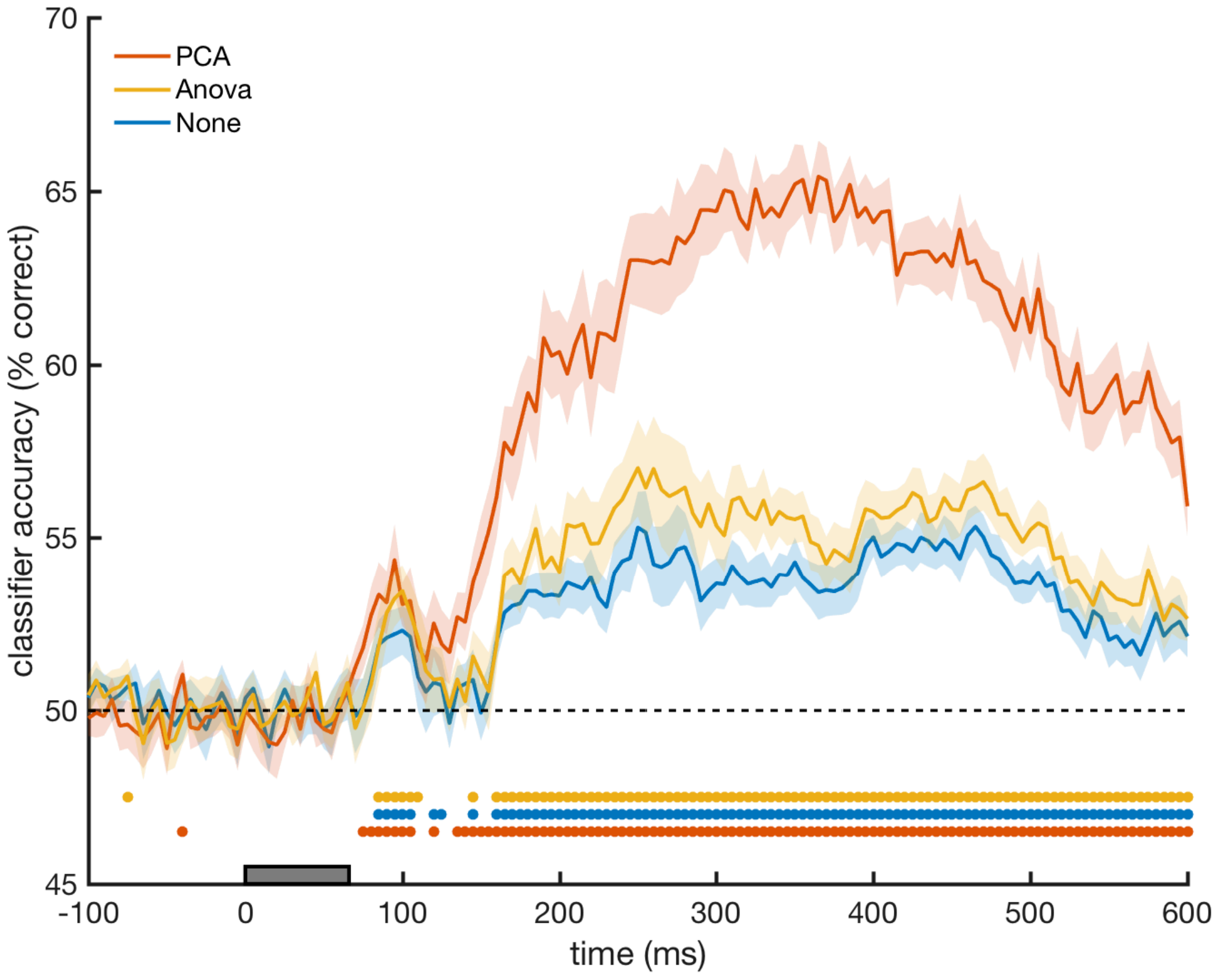

**Figure 6.** The effect of dimensionality reduction methods on decoding performance. The effect of channel selection using ANOVA (yellow line) is marginally better than using the raw data (blue line). Using PCA (red line) yields the largest gain in performance. The shaded area is the standard error across subjects. Discs above the x-axis indicate the time points where decoding performance is significantly higher than chance.

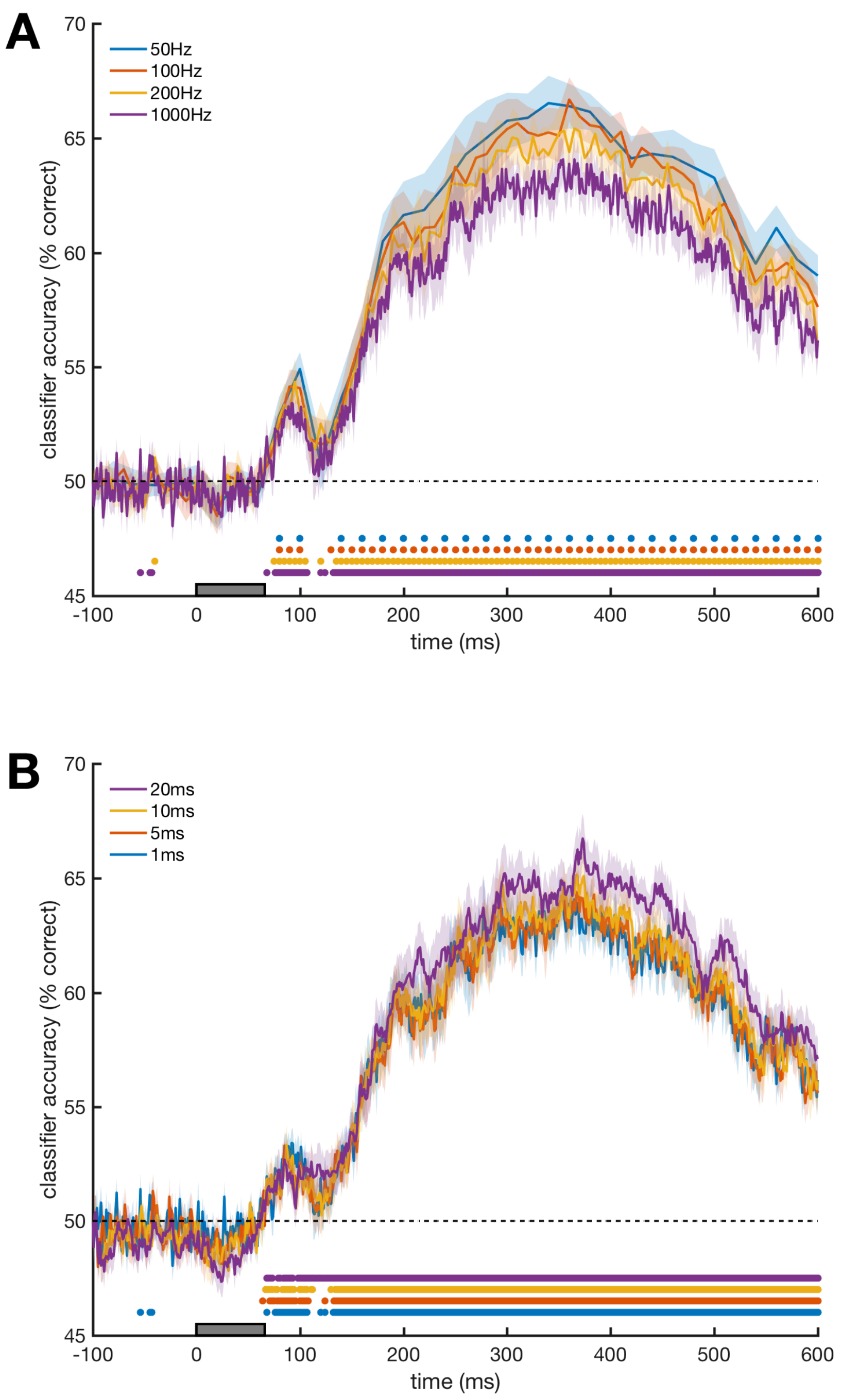

**Figure 7**. The effect of **(A)** subsampling and **(B)** sliding window approaches to improving signal-to-noise on classifier accuracy. The shaded area is the standard error across subjects. Discs above the x-axis indicate the time points where decoding performance is significantly higher than chance.

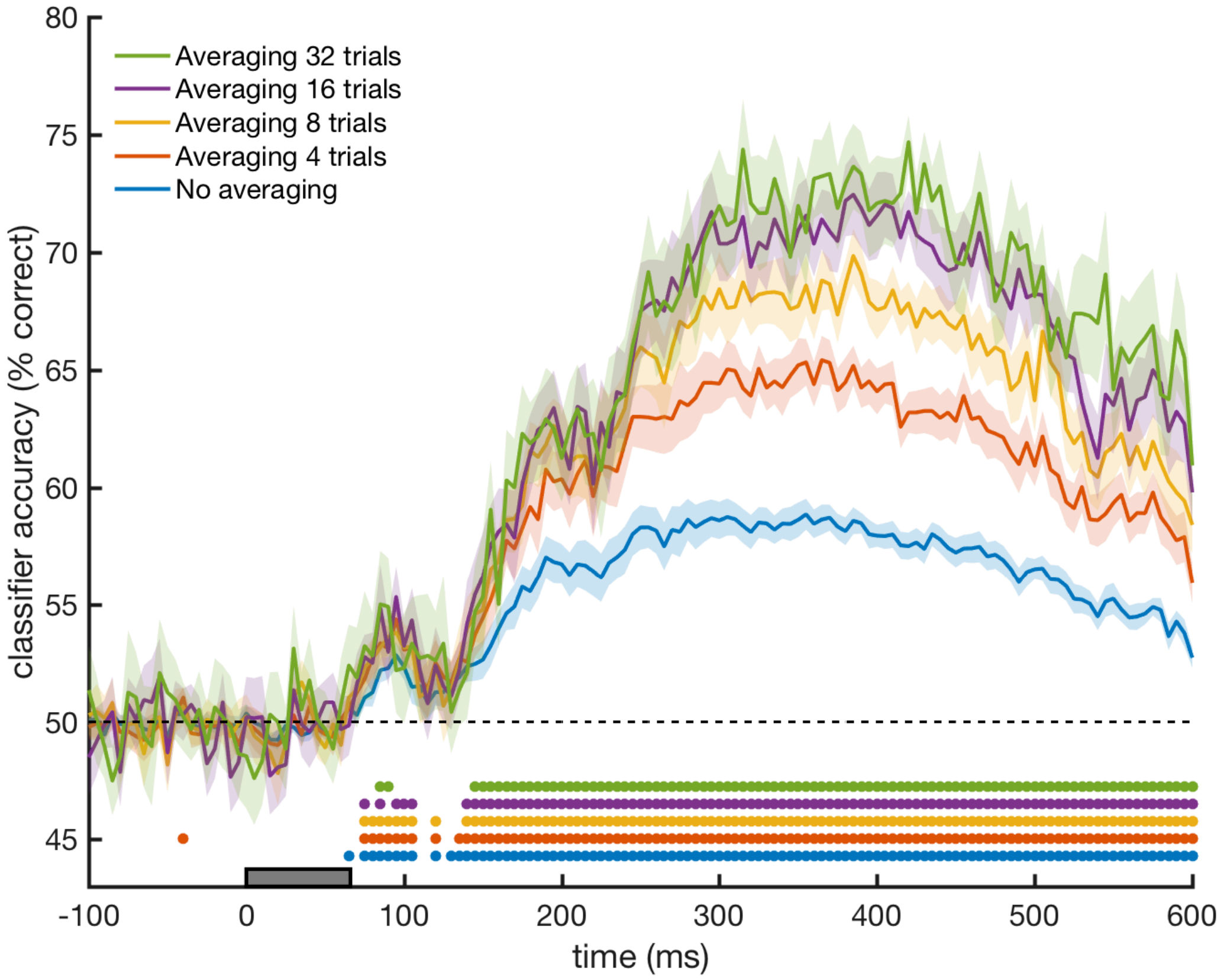

**Figure 8.** The effect of averaging trials on decoding performance. The shaded area is the standard error across subjects. Discs above the x-axis indicate the time points where decoding performance is significantly higher than chance.

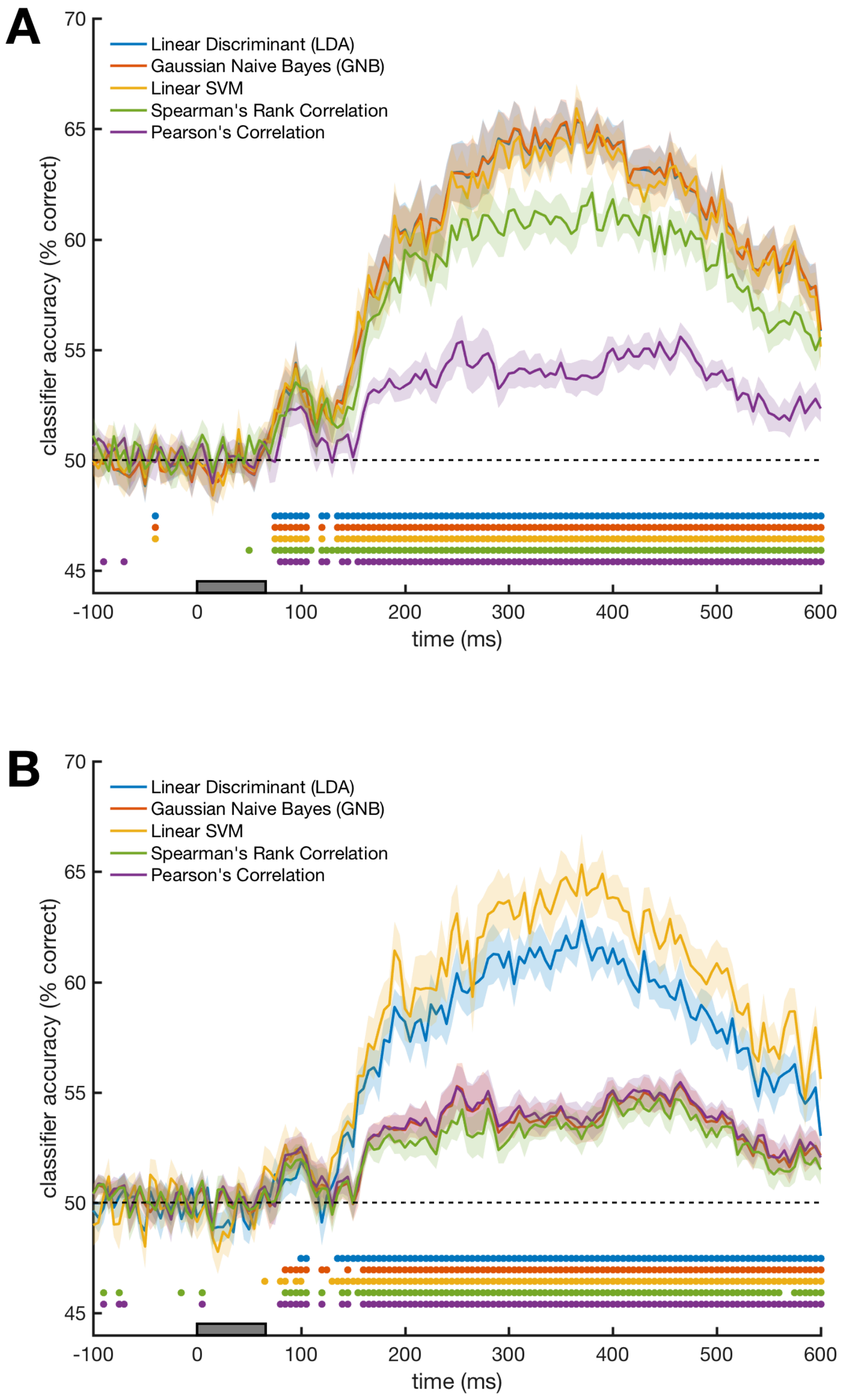

**Figure 9.** Comparison of classification accuracy as a function of classifier type. A. Using the standard decoding pipeline. B. Using the standard pipeline without performing PCA. The shaded area is the standard error across subjects. Discs above the x-axis indicate the time points where decoding performance is significantly higher than chance.

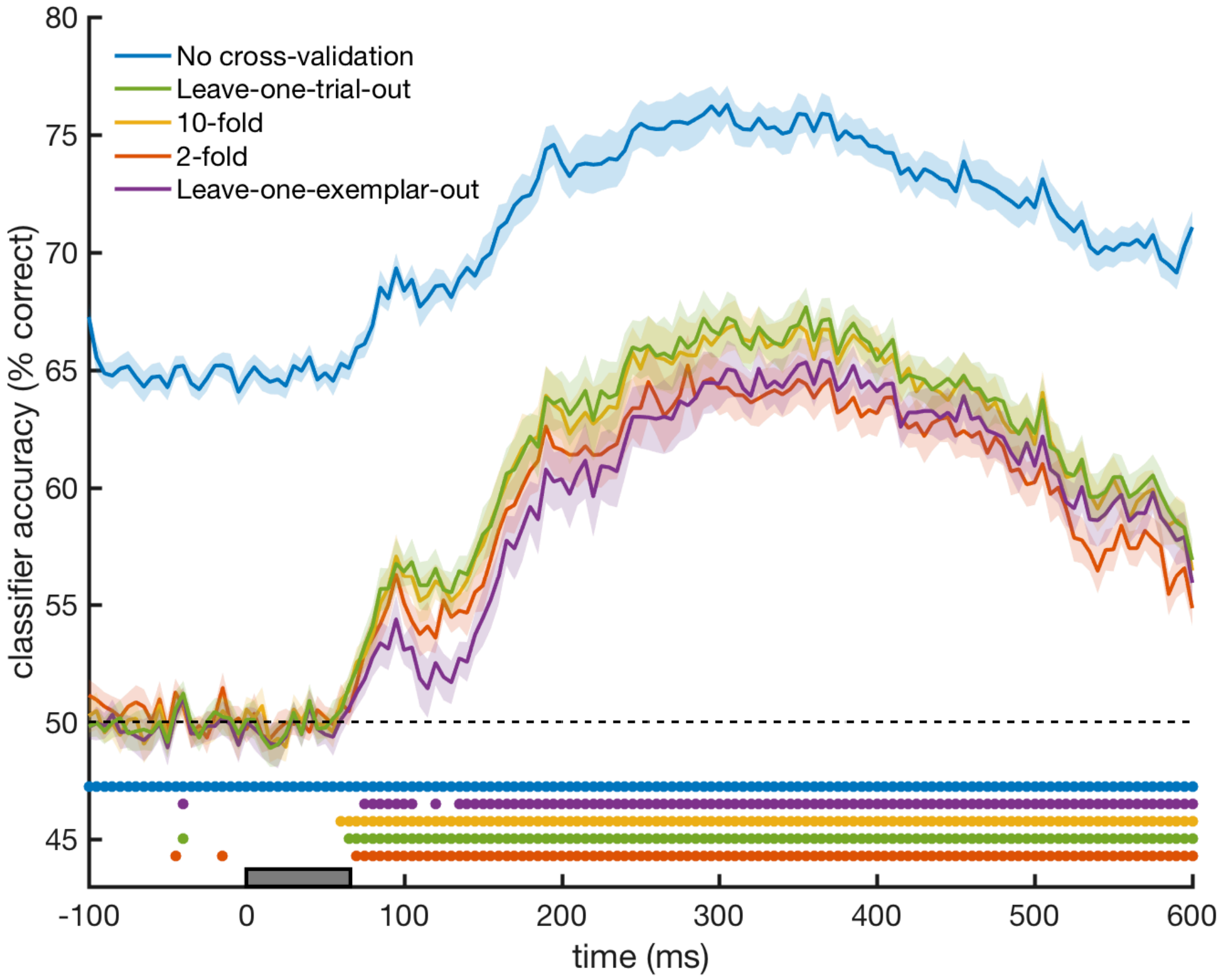

**Figure 10.** Classification accuracy as a function of cross-validation method. The shaded area is the standard error across subjects. Discs above the x-axis indicate the time points where decoding performance is significantly higher than chance.

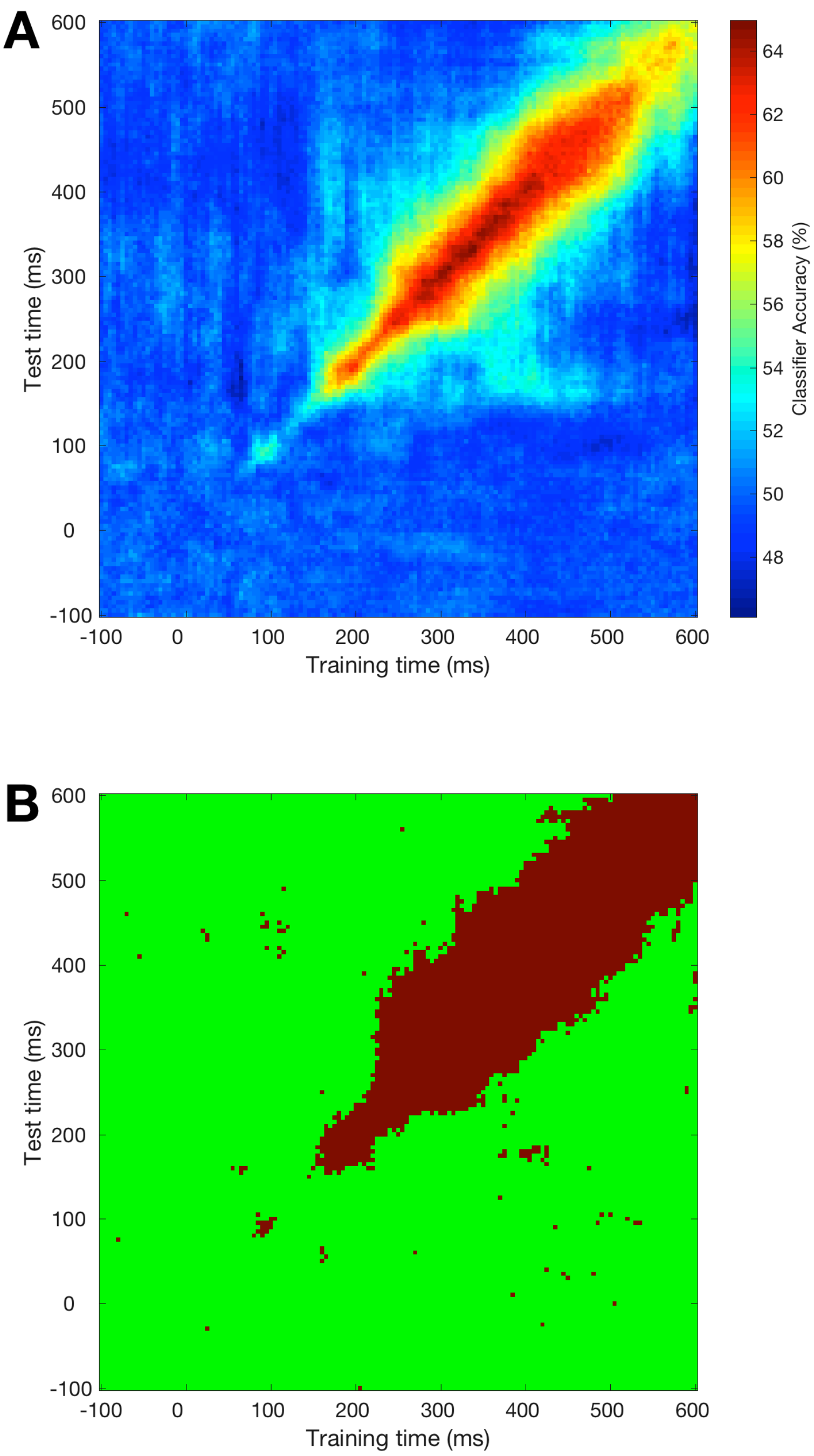

**Figure 11. A.** Temporal generalization of decoding performance. A classifier is trained at one time point, and tested at a different time point. This is repeated for all pairs of time points. The figure shows the generalization accuracy averaged over subjects. B. Map of time point pairs where the generalization was significantly different (red area) from chance (Wilcoxon signed rank test, controlled for multiple comparisons using FDR).

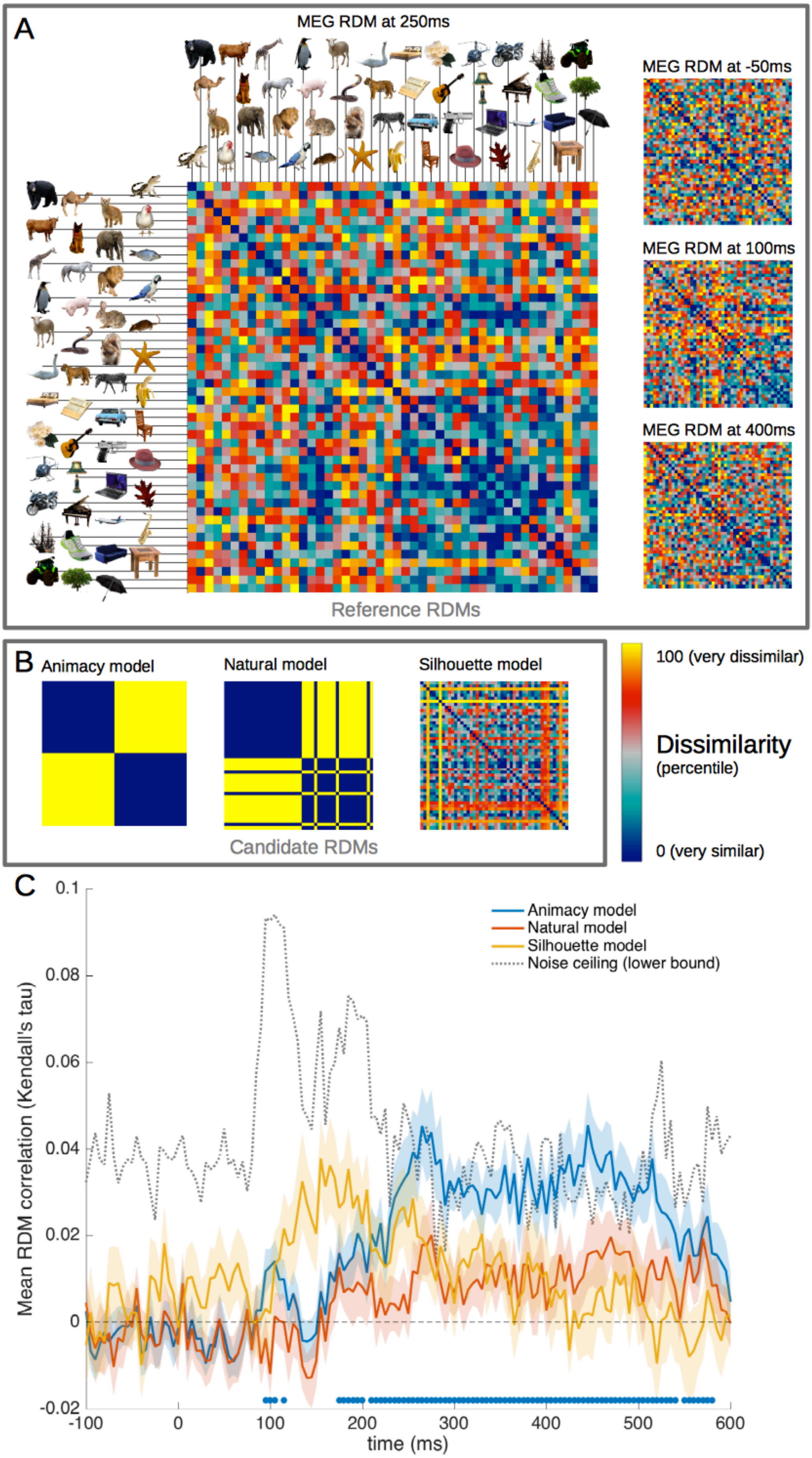

**Figure 12. Model evaluation within the RSA framework. A.** The empirical MEG RDMs averaged across subjects. One cell in the matrix represents the dissimilarity between the MEG activation patterns for one pair of object exemplars. RDMs are shown for four time points: -50ms, 100ms, 250ms, and 400ms. **B.** Three model RDMs, which predict the representational similarity of the brain activation patterns for all object pairs based on different stimulus properties: an Animacy model (Animate vs. Inanimate objects), a Natural model (Natural vs. Artificial objects), and a Silhouette model (based on the visual similarity of the objects' silhouettes). **C.** RSA model evaluation. At each time point, the empirical RDMs for each subject are correlated with the three candidate model RDMs in B. The strength of the average correlations shows how well the candidate models fit the data. Shaded areas represent the standard error over subjects, and the marks above the x-axis indicate time points where the mean correlation was significantly higher than zero (Wilcoxon signed-rank test, controlled for multiple comparisons using FDR). The grey dotted line represents the lower bound of the 'noise ceiling' at each time point, which is the theoretical lower bound of the maximum correlation of any model with the reference RDMs at each time point, given the noise in the data (Nili et al., 2014).

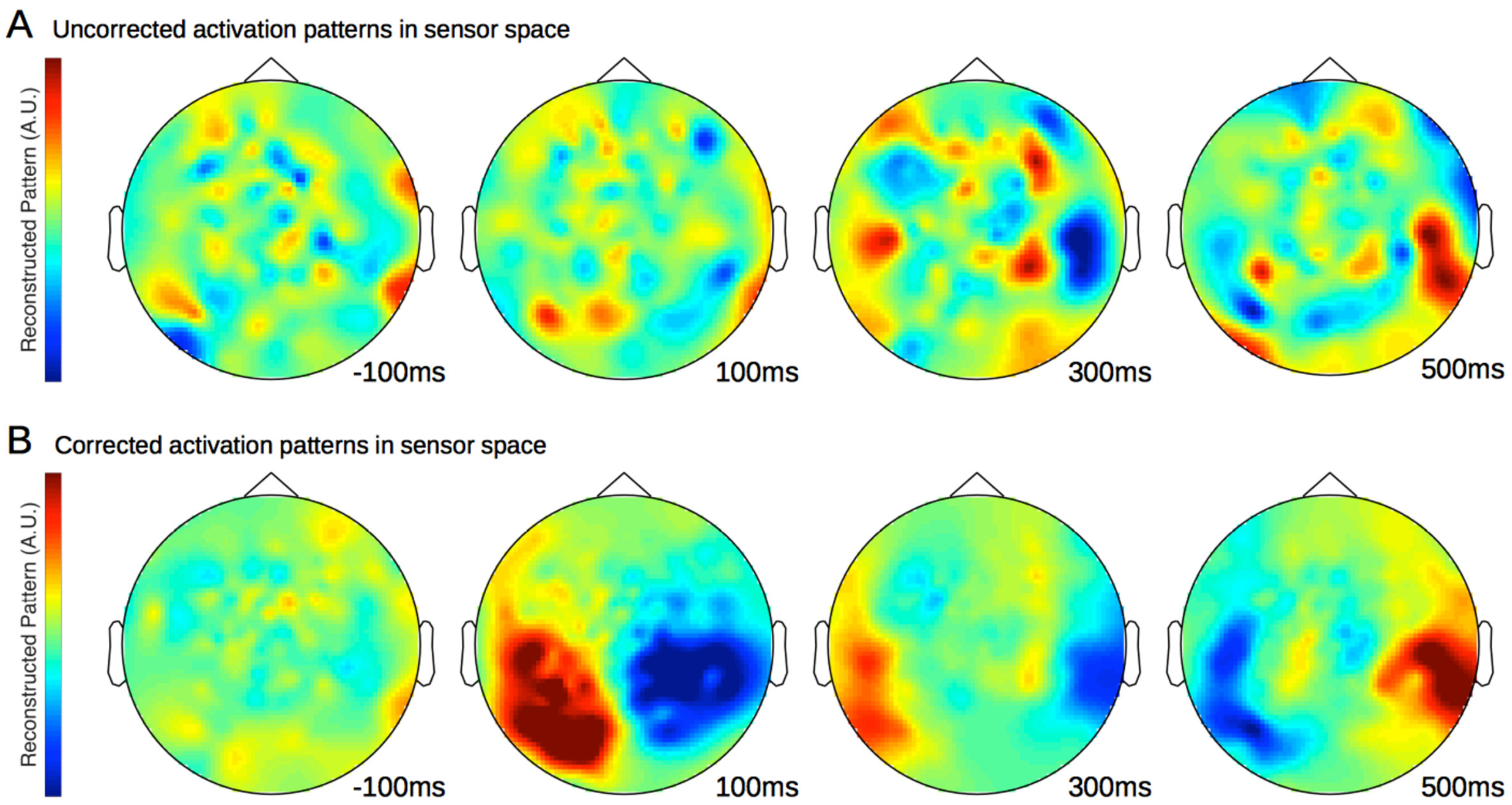

**Figure 13.** Classifier weights projected onto MEG sensor space. The corresponding time points are shown beneath the scalp topographies. Darker colours indicate channels that contribute to animacy decoding. **A.** Uncorrected (raw) weights projections cannot be interpreted directly, as classifiers can assign non-zero weights to channels that contain no class-specific information. **B.** The activation patterns computed from transformed weights (following the method of Haufe et al., (2014)) can be interpreted.

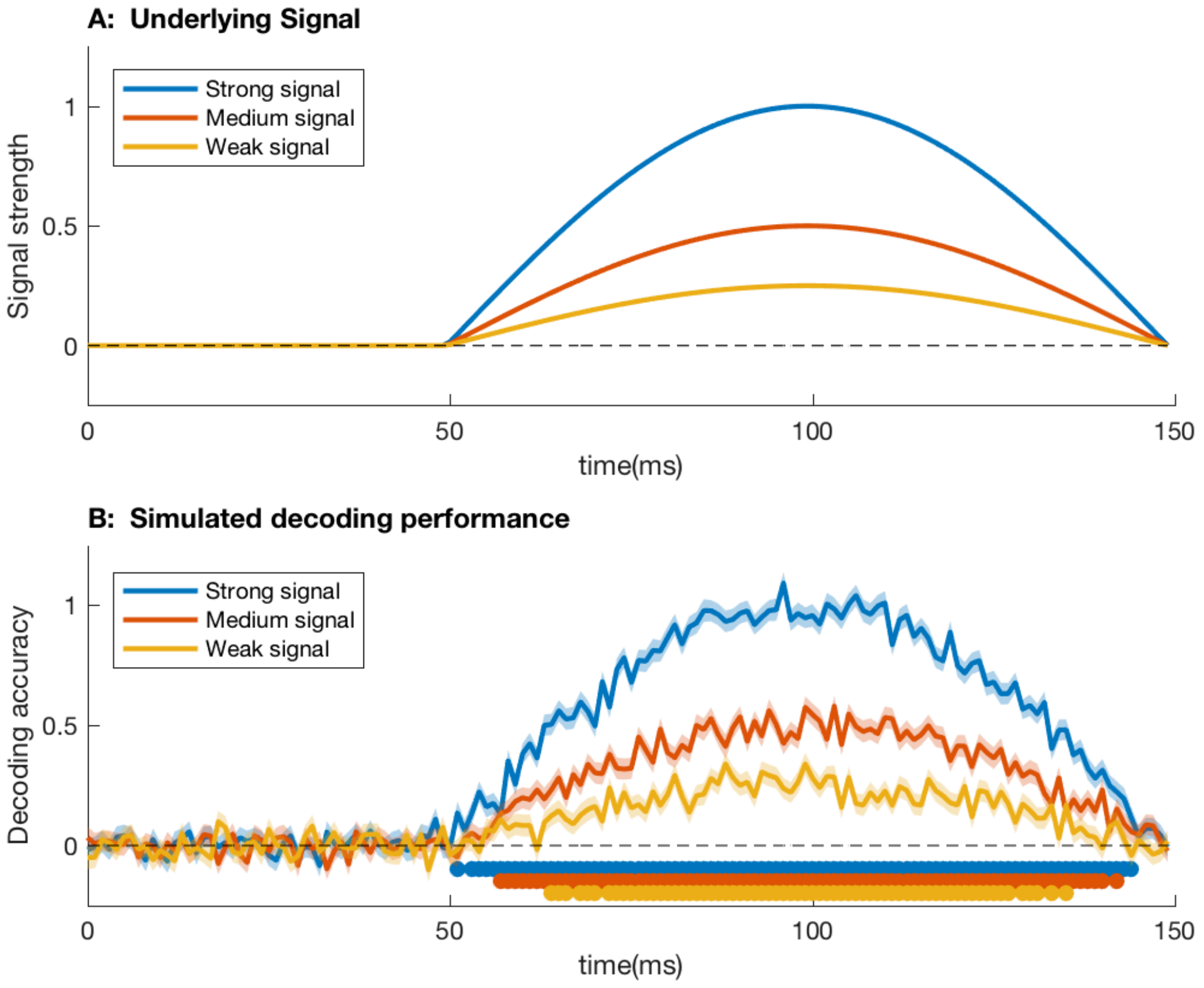

**Figure 14.** Demonstration of how the strength of peak decoding affects decoding onsets using stimulated data. **A.** Three data sets were simulated to have the same onset and peak decoding latencies, but different peak strengths. **B.** Gaussian noise was added to the underlying signals in each set (500 trials per set, $\sigma$=1) and significant decoding (above zero) was assessed across the time-course (signed-rank test, FDR corrected). Coloured discs above the x-axis indicate time points with significant decoding.

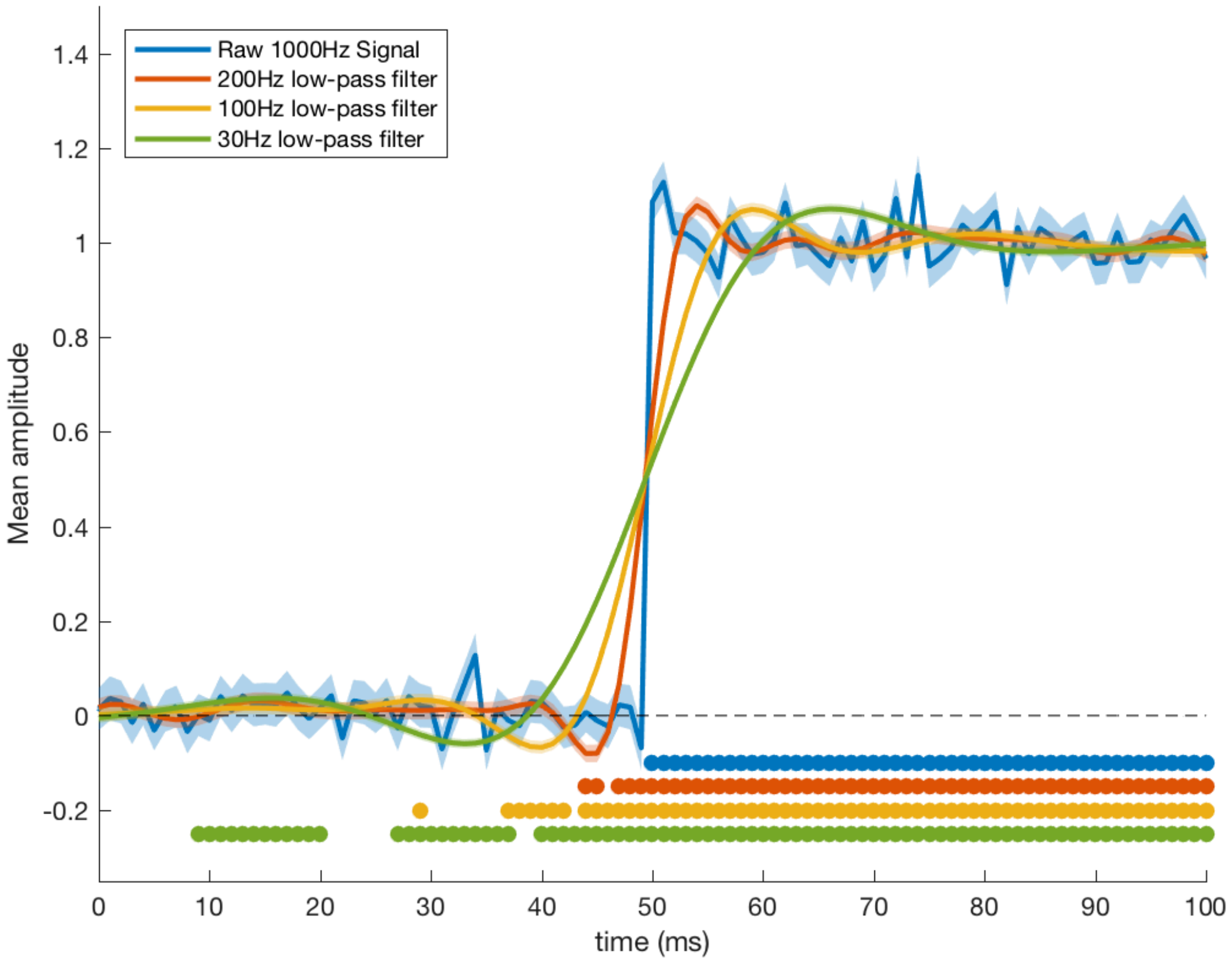

**Figure 15**. The effect of low-pass filtering on decoding onset. In this example, a signal with onset at 50ms was simulated with added Gaussian noise (500 trials, $\boldsymbol{\sigma}$=1). The signal was then low-pass filtered using different cut-off frequencies. Time points where the trial average differed significantly from zero (signed-rank test, FDR corrected) are indicated by the coloured discs above the x-axis.